\documentclass[aps,prl,twocolumn,showpacs,preprintnumbers,
               nofootinbib,float,superscriptaddress]{revtex4-1}

\usepackage{epsfig}
\usepackage{color}
\usepackage[dvipsnames]{xcolor}
\usepackage{float,amsmath}
\usepackage{graphicx}
\usepackage{hyperref}
\usepackage[utf8]{inputenc}
\begin{document}

\title{Hybrid Color Glass Condensate and hydrodynamic description of the\\ Relativistic Heavy Ion Collider small system scan }

\author{Bj\"orn Schenke}
\affiliation{Physics Department, Brookhaven National Laboratory, Upton, NY 11973, USA}

\author{Chun Shen}
\affiliation{Department of Physics and Astronomy, Wayne State University, Detroit, Michigan 48201, USA}
\affiliation{RIKEN BNL Research Center, Brookhaven National Laboratory, Upton, NY 11973, USA}

\author{Prithwish Tribedy}
\affiliation{Physics Department, Brookhaven National Laboratory, Upton, NY 11973, USA}

\begin{abstract}
Multi-particle correlation observables in the Relativistic Heavy Ion Collider small system scan are computed in a framework that contains both initial state momentum anisotropies from the Color Glass Condensate effective theory and final state hydrodynamic evolution. The initial state is computed using the IP-Glasma model and coupled to viscous relativistic hydrodynamic simulations, which are followed by microscopic hadronic transport. All parameters of the calculation were previously constrained using experimental data on Au+Au collisions at the same center of mass energy. We find that the qualitative features of the experimental data, such as the system and centrality dependence of the charged hadron momentum anisotropy, can only be reproduced when final state interactions are present.
On the other hand, we also demonstrate that the details of the initial state are crucially important for the quantitative description of observables in the studied small systems, as neglecting the initial transverse flow profile or the initial shear stress tensor, which contain information on the momentum anisotropy from the Color Glass Condensate, has dramatic effects on the produced final state anisotropy. We further show that the initial state momentum anisotropy is correlated with the observed elliptic flow in all small systems, with the effect increasing with decreasing multiplicity. We identify the precise measurement of $v_2$ in d+Au and Au+Au collisions at RHIC energy at the same multiplicity as a means to reveal effects of the initial state momentum anisotropy.
\end{abstract}

\maketitle


The origins of azimuthal anisotropies in the produced hadron momentum distributions observed in high energy collisions involving protons or other small nuclei at the Relativistic Heavy Ion Collider (RHIC) and the Large Hadron Collider (LHC) have been under strong debate \cite{Dusling:2015gta,Schlichting:2016kjw}. The main question is whether initial state momentum correlations, which originate from both classical and quantum effects in the multi-particle production process \cite{Dumitru:2008wn,Kovner:2010xk,Dumitru:2010iy,Kovner:2011pe,Dusling:2012iga,Levin:2011fb,Dusling:2012wy,Dusling:2013qoz,Dumitru:2014dra,Dumitru:2014yza,Schenke:2015aqa,McLerran:2015sva,Schenke:2016lrs,Dusling:2017dqg,Dusling:2017aot,Mace:2018vwq,Mace:2018yvl,Kovner:2018fxj} and can be computed using Color Glass Condensate (CGC) effective theory \cite{McLerran:1994ni,McLerran:1994ka,Iancu:2003xm}, contribute significantly to the observed anisotropy coefficients, or whether final state interactions in the produced medium \cite{Bozek:2011if,Bozek:2012gr,Bozek:2013df,Bozek:2013uha,Bozek:2013ska,Bzdak:2013zma,Qin:2013bha,Werner:2013ipa,Kozlov:2014fqa,Schenke:2014zha,Romatschke:2015gxa,Shen:2016zpp,Weller:2017tsr,Mantysaari:2017cni} are absolutely dominant. 

Both contributions have been studied in a combined framework of the IP-Glasma initial state \cite{Schenke:2012wb,Schenke:2012hg} and a microscopic description of the gluonic final state interactions via solutions of the Boltzmann equation in \cite{Greif:2017bnr}. It was found that initial state correlations survive for transverse momenta $p_T\gtrsim 1.5\,{\rm GeV}$ even in high multiplicity p+Pb events, with their effect and affected $p_T$ range increasing with decreasing multiplicity. We go beyond this study in two essential ways. First, we allow for the comparison to experimental data by computing hadronic final states instead of studying parton spectra, and second, we quantify how strongly the CGC initial state momentum anisotropy can affect experimental observables if final state interactions are described by realistic hydrodynamic simulations that were constrained by heavy ion data.

Previous calculations involving the IP-Glasma initial state coupled to a hydrodynamic description of the final state evolution \cite{Schenke:2010nt,Schenke:2010rr,Schenke:2011bn} in principle also contain information on both CGC initial state and geometry driven final state sources of correlations. However, only the work on p+Pb collisions at LHC in \cite{Mantysaari:2017cni} included the full initial energy momentum tensor from the IP-Glasma calculation to initialize hydrodynamics, which is necessary in order to keep the full initial state information, and to conserve energy and momentum when switching descriptions. 

The striking results from the recent small system scan at RHIC \cite{PHENIX:2018lia} provide a strong motivation to extend our work to lower energies and to analyze in more detail the features of initial and final state effects on the observed momentum anisotropies. We study different collision systems (p+p, p+Au, d+Au, and $^3$He+Au) at the top RHIC energy, using the hybrid framework of IP-Glasma + \textsc{Music} hydrodynamics + UrQMD \cite{Bass:1998ca,Bleicher:1999xi} with parameters constrained by heavy ion collisions \cite{Schenke:2019ruo}. Apart from presenting results for two- and four- particle correlations and comparison to experimental data, we show the multiplicity and system dependence of the initial elliptic momentum anisotropy, and its correlation with the observable elliptic flow. 

The energy momentum tensor from the IP-Glasma framework, used to initialize the hydrodynamic simulation, includes the single particle CGC momentum anisotropy. We study individually the effects of the initial transverse flow profile and the viscous components of the energy momentum tensor. This allows us to determine the importance of different features of the initial state description, many of which are neglected in recently explored approximations to the IP-Glasma model \cite{Nagle:2018ybc,Lim:2018huo,Sievert:2019zjr} (see also \cite{Moreland:2014oya}). 

\paragraph{Framework.}
We employ the hybrid framework described in \cite{Mantysaari:2017cni} and \cite{Schenke:2019ruo}, consisting of the boost-invariant IP-Glasma initial state, which provides the spatial ($\mathbf{x}_\perp$) distribution of (locally anisotropic) energy-momentum tensors as input for the hydrodynamic simulation \textsc{Music}, which in turn is coupled to the hadronic transport simulation UrQMD, which describes the low energy density regime.

The IP-Glasma model samples nucleon positions from nuclear density distributions, including a three-hotspot substructure for each nucleon to generate the nuclear geometry of a single configuration. For the deuteron, we sample a Hulthen wave function \cite{Hulthen:1957,Miller:2007ri}, for $^3$He we use the same configurations as in \cite{Nagle:2013lja}, obtained using Green’s function Monte Carlo calculations \cite{Carlson:1997qn}. For gold nuclei we use Woods-Saxon distributions as described in \cite{Schenke:2019ruo}. Given that geometry, the IPSat model \cite{Kowalski:2003hm}, constrained by HERA data \cite{Rezaeian:2012ji} provides the distribution of the color charge density, which determines the variance of the Gaussian distributed color charges, which are then sampled. These color charges are the sources for the gluon fields, which are found by solving the SU(3) Yang-Mills equations numerically \cite{Krasnitz:1998ns}. From the two incoming sheets of gluon fields one determines the gluon fields after the collision \cite{Krasnitz:1998ns,Schenke:2012wb,Schenke:2012hg}, which are then evolved in time. At time $\tau_{\rm init}$ we switch from the Yang-Mills to the hydrodynamic description by using the gluon fields' energy momentum tensor $T^{\mu\nu}_{\rm CYM}(\mathbf{x}_\perp)$ as initial condition for the hydrodynamic simulations. We use $\tau_{\rm init}=0.4\,{\rm fm}/c$ and will comment on the  effect of varying $\tau_{\rm init}$ between $0.2$ and $0.6\,{\rm fm}/c$.

As in \cite{Mantysaari:2017cni} and \cite{Schenke:2019ruo}, we include the full energy momentum tensor, which has the shear stress contribution
\begin{equation}
\pi^{\mu\nu}=T^{\mu\nu}_{\rm CYM}-\frac{4}{3}\varepsilon u^\mu u^\nu + \frac{\varepsilon}{3}g^{\mu\nu}\,.
\end{equation}
The initial energy density $\varepsilon$ and flow velocity $u^\mu$ can be extracted by solving $u_\mu T^{\mu\nu}_{\rm CYM}=\varepsilon u^\nu$. 

 On the Yang-Mills side, the pressure is given by $P_{\rm CYM}=\varepsilon/3$, while in the hydrodynamic simulation we employ a lattice QCD equation of state \cite{Bazavov:2014pvz,Moreland:2015dvc} with pressure $P_{\rm lat}$. There are various ways of dealing with this mismatch. One way is to absorb the difference in an effective initial bulk viscous term $\Pi = \varepsilon/3-P_{\rm lat}$, which vanishes on a time scale approximately given by the bulk relaxation time. The initial $\Pi$ constructed this way leads to an additional outward push \cite{Schenke:2019ruo}. Alternatively, one can accept a jump in the pressure at the time of matching, by setting the initial $\Pi=0$. The former option is our preferred one, as it was used in \cite{Schenke:2019ruo}. We will also study the latter option, which will determine one contribution to our systematic uncertainties.

Once initialized at time $\tau_{\rm init}$, the energy momentum tensor follows $\partial_\mu T^{\mu\nu} = 0$, where
\begin{equation}
T^{\mu\nu} = \varepsilon u^\mu u^\nu - (P + \Pi) \Delta^{\mu\nu} + \pi^{\mu\nu}\,,
\end{equation}
with $\Delta^{\mu\nu} = g^{\mu\nu}-u^\mu u^\nu$. We use the same second order constitutive relations for the shear and bulk viscous parts as derived in \cite{Denicol:2012cn,Denicol:2014vaa} and used in \cite{Schenke:2019ruo}, and solve them within the simulation \textsc{Music} \cite{Schenke:2010nt,Schenke:2010rr,Schenke:2011bn}. 
We use the same shear and bulk viscosities as in \cite{Schenke:2019ruo}. Because the second order transport coefficients are not well known, we will vary all (except the relaxation times) between what was used in \cite{Schenke:2019ruo} (values determined from a Boltzmann gas in the small mass limit \cite{Denicol:2012cn,Denicol:2014vaa}) and zero, which constitutes the other contribution to our systematic errors.

When the medium temperature drops to the switching temperature $T_{\rm sw}=145\,{\rm MeV}$, the fluid is converted to particles by first computing the particle spectra according to the Cooper-Frye formula \cite{Cooper:1974mv}, using equilibrium distributions $f_0$ with viscous corrections $\delta f$, given in  \cite{Dusling:2009df,Bozek:2009dw,Paquet:2015lta} for shear and bulk viscous terms. From these non-equilibrium distribution functions $f=f_0+\delta f$, we sample particles on the switching surface\footnote{We employ the publicly available numerical code iSS:\\ \url{https://github.com/chunshen1987/iSS}} that then undergo the microscopic transport processes of UrQMD \cite{Bass:1998ca,Bleicher:1999xi}. To ensure enough statistics, each individual hydrodynamic hyper-surface is oversampled until we reach at least 100,000 particles per unit of rapidity.

\begin{figure}[tb]
  \includegraphics[width=0.47\textwidth]{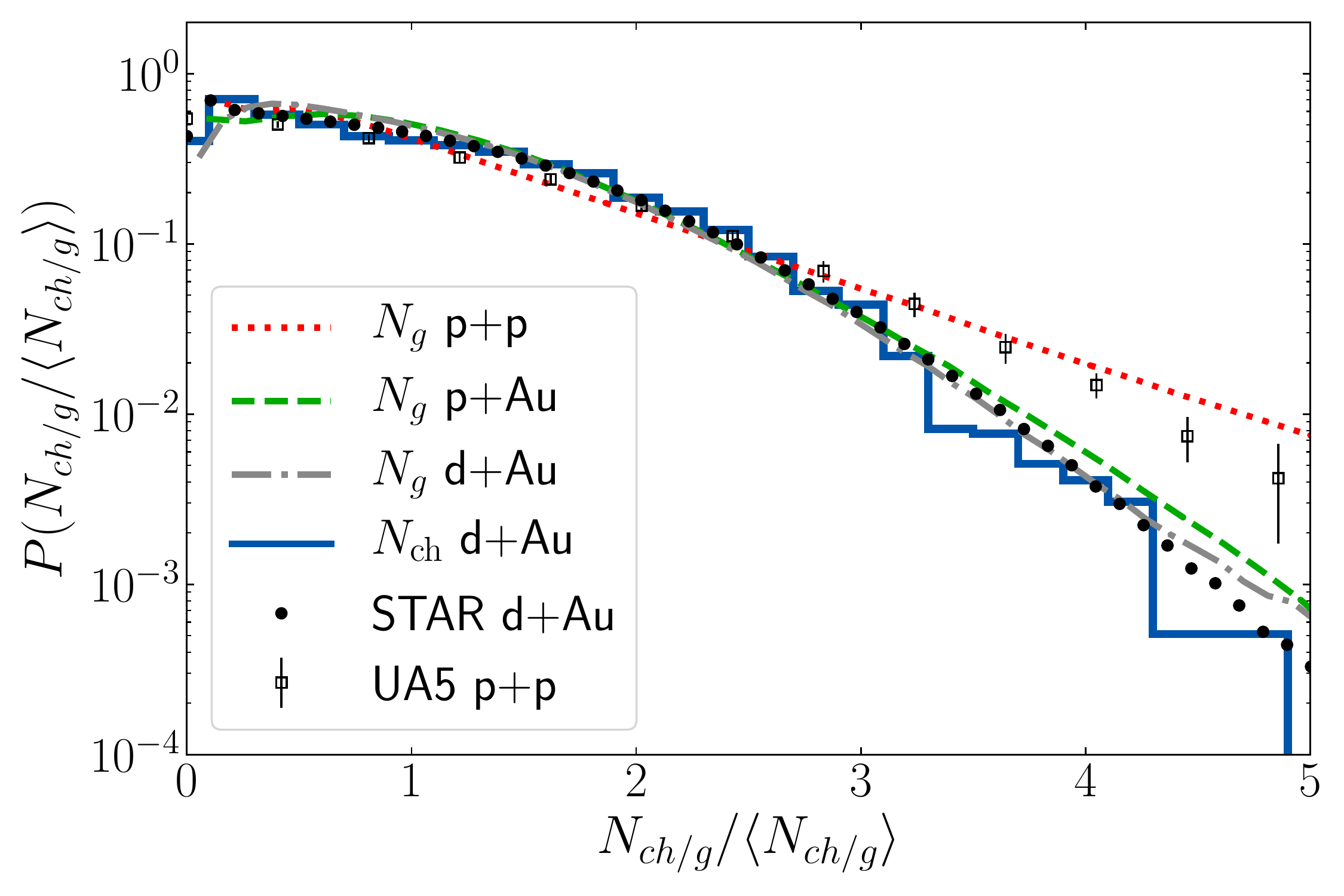}
  \caption{Gluon multiplicity $N_g$ distributions from the IP-Glasma model in p+p, p+Au, and d+Au collisions and charged particle multiplicity $N_{\rm ch}$ distribution after hydrodynamic evolution in d+Au collisions, scaled by the mean multiplicity and compared to scaled $N_{\rm ch}$ distributions from STAR \cite{Abelev:2008ab} for d+Au (uncorrected) and UA5 \cite{Ansorge1989} for p+p (corrected). \label{fig:multDist}}
\end{figure}

\paragraph{Multiplicity distributions.}
Multi-particle correlations in small systems are very sensitive to event-by-event fluctuations. Consequently, basic observables like the charged particle multiplicity distributions must be described correctly. We present both the gluon multiplicity distributions (folded with a Poisson distribution to estimate the effect of (grand canonical) sampling on the switching surface), which is used to determine our centrality classes, and for d+Au collisions also the final state charged hadron distribution.

Results for multiplicity distributions scaled by the mean multiplicity in p+p, p+Au, and d+Au systems at $\sqrt{s}=200\,{\rm GeV}$ are shown in Fig.\,\ref{fig:multDist}, comparing to experimental data (of the uncorrected charged particle multiplicity on the Au going side) in d+Au collisions from the STAR Collaboration \cite{Abelev:2008ab} and (to the corrected charged particle multiplicity data within $|\eta|<0.5$) in p+p collisions from the UA5 Collaboration \cite{Ansorge1989}. While the statistics for the computed charged hadron multiplicity distribution is limited, one can see that the (scaled) gluon distribution is a good proxy for the final state charged hadron distribution. Furthermore, the shape of the experimental distributions in p+p and d+Au collisions is well reproduced in our framework.

\begin{figure}[tb]
  \includegraphics[width=0.47\textwidth]{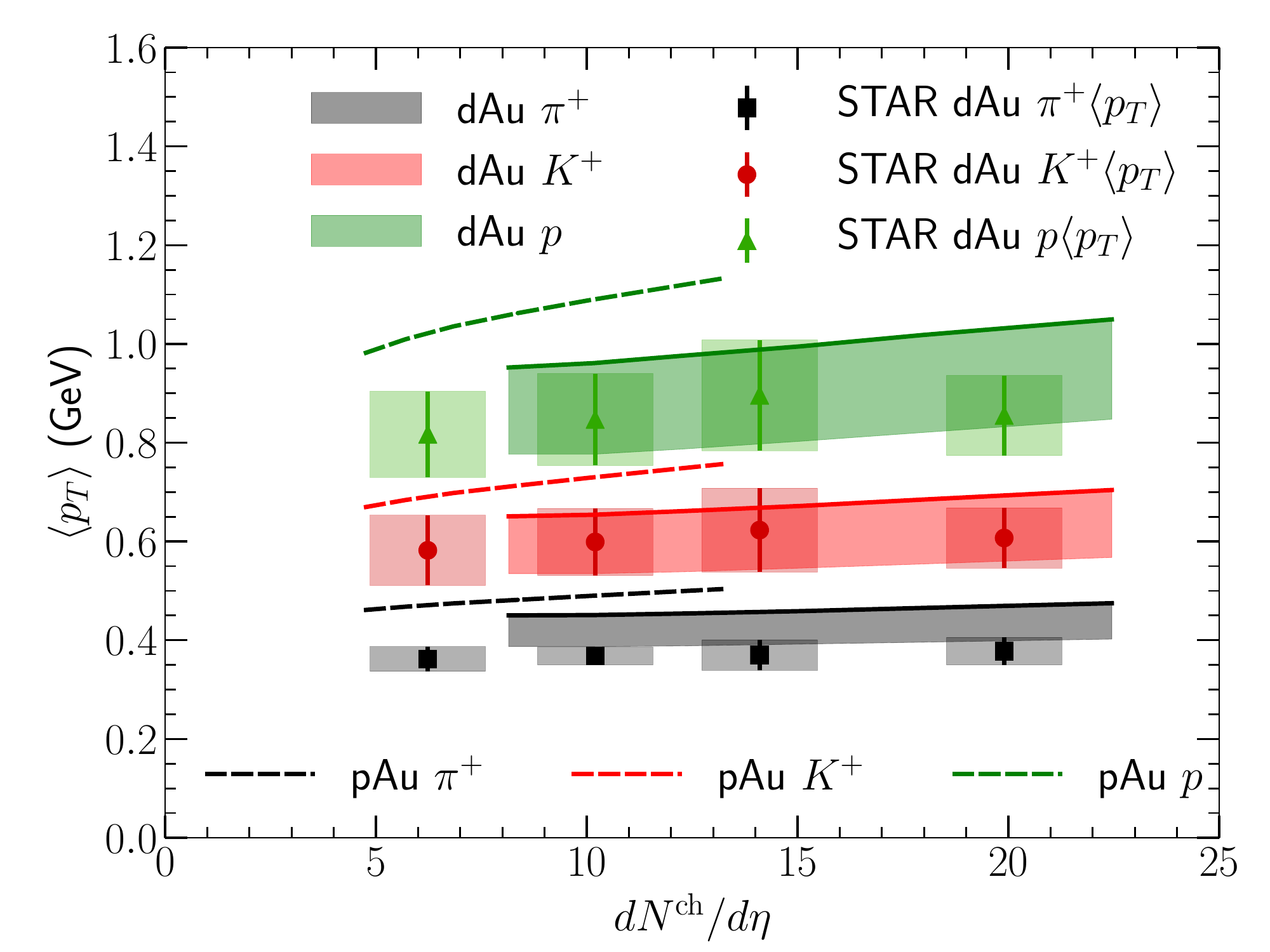}
  \caption{The $\langle p_T\rangle$ of $\pi^+$, $K^+$, and $p$ as a function of charged particle multiplicity in d+Au collisions compared to experimental data from the STAR Collaboration \cite{Abelev:2008ab}. Dashed lines show $\langle p_T \rangle$ in p+Au collisions. See text for details. \label{fig:mpt}}
\end{figure}

\begin{figure*}[htb]
  \includegraphics[width=0.85\textwidth]{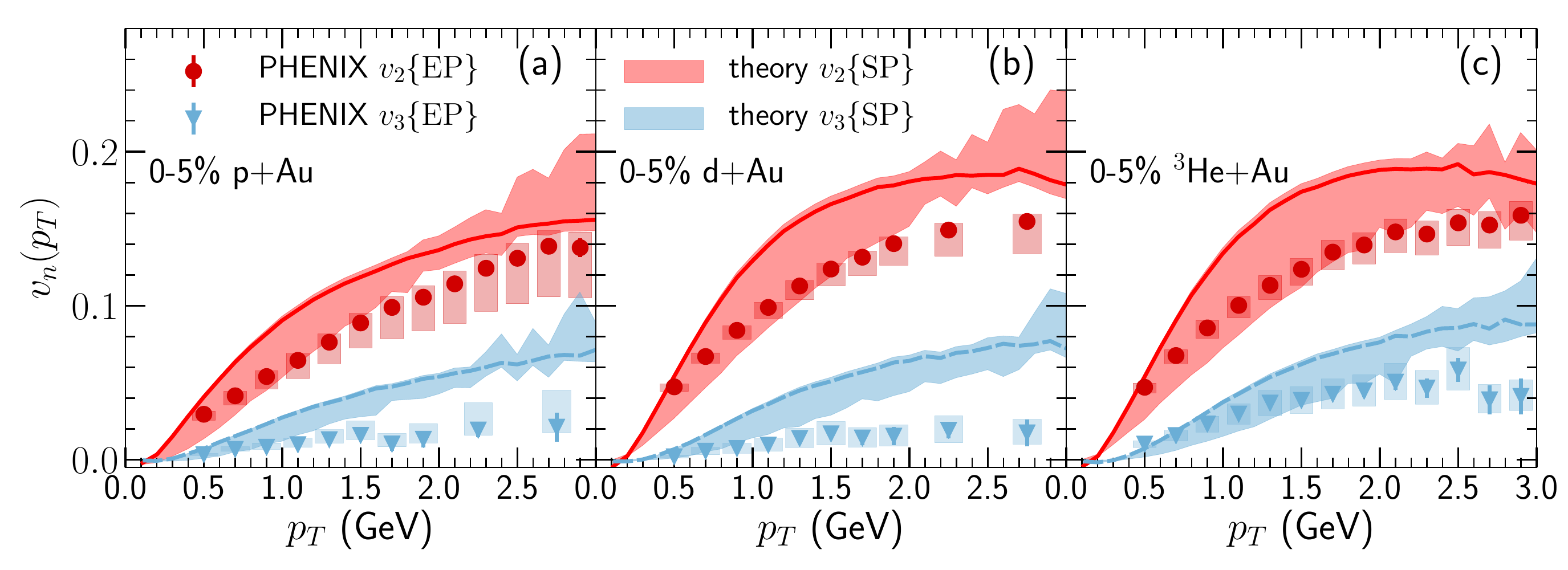} \vspace{-0.4cm}
  \caption{Comparison of the computed $v_n\{\mathrm{SP}\}(p_T)$ to the experimental results on $v_n\{\mathrm{EP}\}(p_T)$ from the PHENIX Collaboration \cite{PHENIX:2018lia} for p+Au (a), d+Au (b), $^3$He+Au (c), each in the respective 0-5\% centrality class. See text for details. \label{fig:vn_pAu_dAu}}
\end{figure*}

\begin{figure}[ht]
\begin{minipage}[t]{0.45\textwidth} 
  \includegraphics[width=\textwidth]{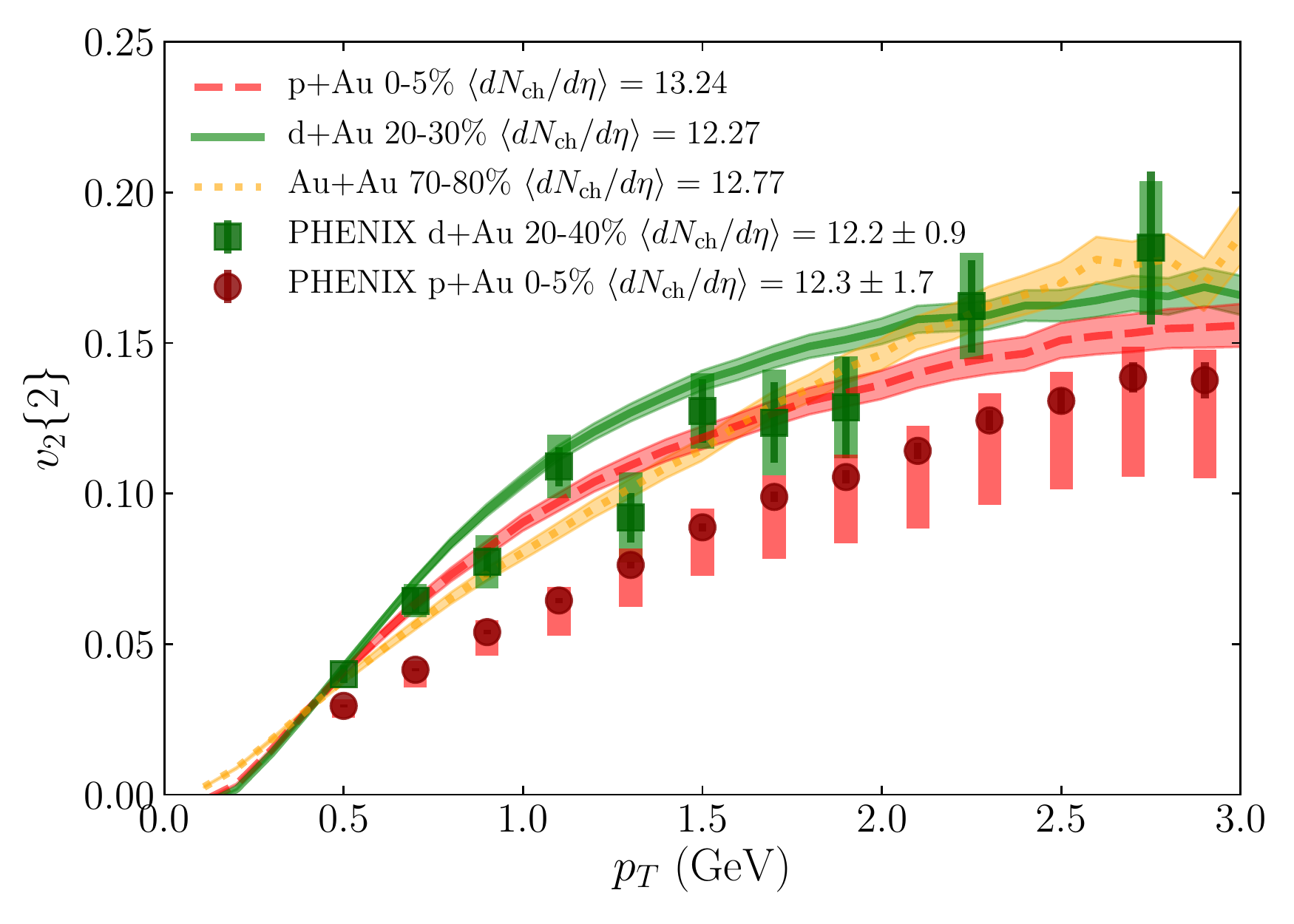}
  \caption{Comparison of the $v_2\{\mathrm{SP}\}(p_T)$ for three different systems at approximately the same multiplicity. \label{fig:v2pT}}
   \vspace{0.1cm}
  \includegraphics[width=\textwidth]{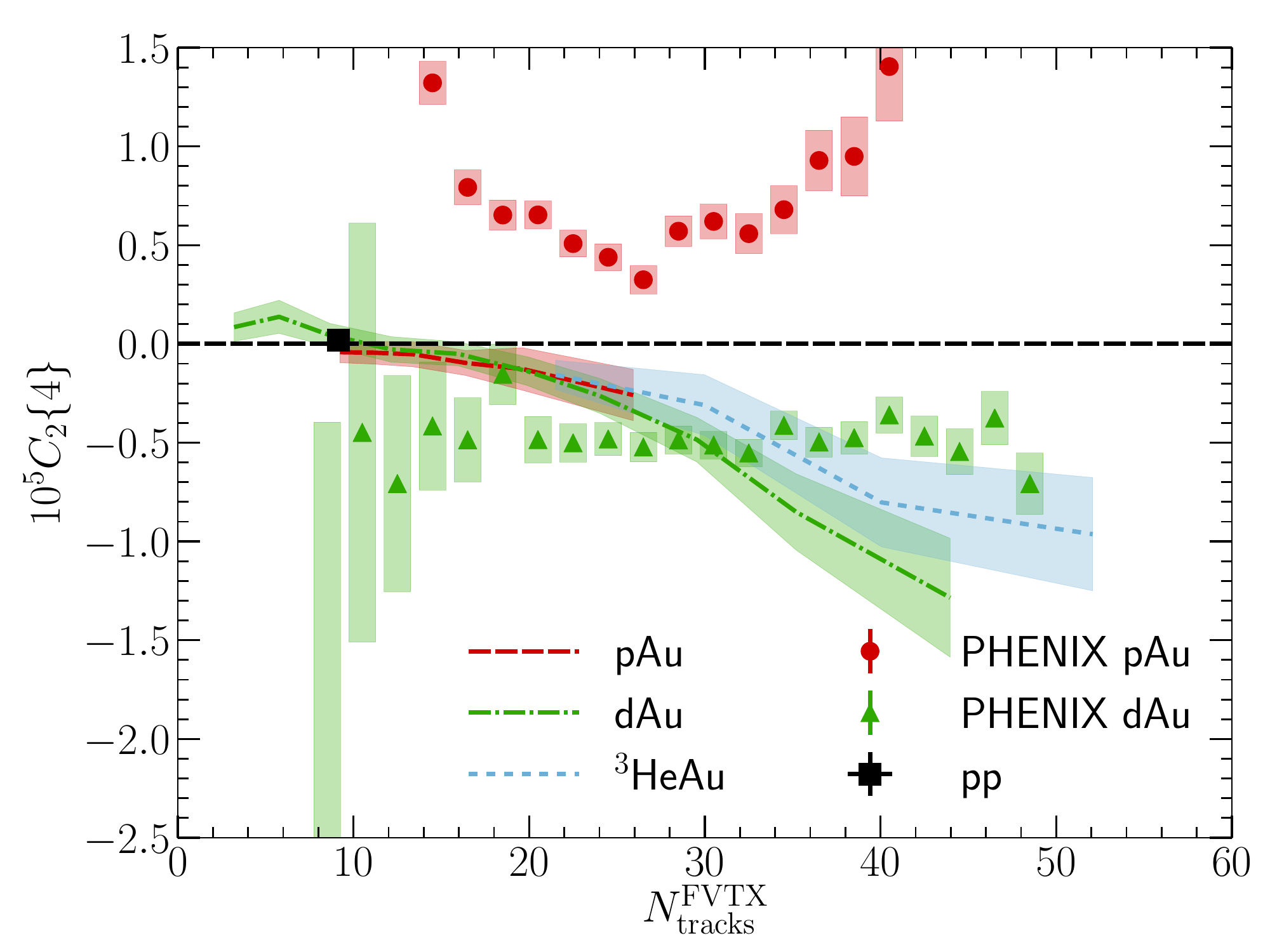}
  \caption{Comparison of $c_2\{4\}$ in p+p, p+Au (dashed), d+Au (dash-dotted), and $^3$He+Au (short dashed) collisions with experimental d+Au data from the PHENIX Collaboration \cite{Aidala:2017ajz}.\label{fig:c24} }
   \vspace{-1cm}
\end{minipage}
\end{figure}

\paragraph{Mean transverse momentum.}
Being sensitive to the initial system size, equation of state, and the bulk viscosity, the mean transverse momentum $\langle p_T\rangle$ is an important observable to constrain models of heavy ion (and smaller system) collisions. In Fig.\,\ref{fig:mpt} we compare our results for $\langle p_T\rangle$ of pions, kaons, and protons in d+Au collisions as a function of charged particle multiplicity to experimental data from the STAR Collaboration \cite{Abelev:2008ab}. The line corresponds to the same parameter set used to describe Au+Au collisions in \cite{Schenke:2019ruo}. The lower edge of the band results from instead setting the initial $\Pi$ and second order transport coefficients (except for the relaxation times) to zero. Agreement with the experimental data is good for kaons and protons, while pions are overestimated, especially when taking into account initial $\Pi$ and second order coefficients. We show results for $\langle p_T \rangle$ obtained in p+Au collisions for comparison. The generally more compact p+Au collisions produce a larger $\langle p_T\rangle$ at the same multiplicity.

\paragraph{Azimuthal anisotropies.}
We determine the transverse momentum dependent elliptic and triangular anisotropy from two-particle correlations and compare to experimental data from the PHENIX Collaboration for p+Au, d+Au, and $^3$He+Au collisions at $\sqrt{s}=200\,{\rm GeV}$ \cite{PHENIX:2018lia}. The experimental $v_2(p_T)$ is obtained using an event-plane method with the second and third order event-planes determined at forward rapidities in the Au going direction, 
and using charged hadrons in the mid-rapidity region.
Our calculation is boost-invariant and uses the scalar product method, which should best approximate the event-plane method used in the experiment when the event-plane resolution is small~\cite{Luzum:2012da}.

We show our results for all three systems in their respective 0-5\% centrality bins and compare to data from the PHENIX Collaboration in Fig.\,\ref{fig:vn_pAu_dAu}. The bands again result from varying the initial $\Pi$ and second order transport coefficients, with the trends being the same as in the case of $\langle p_T\rangle$ for $p_T<2\,{\rm GeV}$. The line is for the parameters constrained in Au+Au collisions \cite{Schenke:2019ruo}. Because of the different spatial geometries of the three systems $\varepsilon_2^{^3{\rm He+Au}} \simeq \varepsilon_2^{\rm d+Au} > \varepsilon_2^{\rm p+Au}$ and $\varepsilon_3^{^3{\rm He+Au}} > \varepsilon_3^{\rm d+Au} \simeq \varepsilon_3^{\rm p+Au}$ \cite{Nagle:2013lja,PHENIX:2018lia} one expects a corresponding ordering of the final state azimuthal momentum anisotropies, if final state response to the geometry is sizable.
Indeed, we observe some of the expected ordering of $v_2(p_T)$ between systems, especially at larger $p_T$. The difference in multiplicity in the three 0-5\% centrality classes also plays a role for this result. The computed $v_3$ is generally greater than the experimental data, except at the lowest transverse momenta. The difference between the p/d+Au $v_3$ and the one in $^3$He+Au collisions is significantly smaller than in the experimental result.

In Fig.\,\ref{fig:v2pT} we demonstrate that $v_2(p_T)$ also shows the ordering expected from the geometry when selecting events with the same multiplicity. We also found that when using smoother nucleons (than the ones used and constrained in \cite{Mantysaari:2016ykx,Mantysaari:2016jaz}) the difference between d+Au $v_2$ and p+Au $v_2$ at the same multiplicity is increased, highlighting a sensitivity to sub-nucleonic fluctuations. Fig.\,\ref{fig:v2pT} also shows a comparison to PHENIX data, indicating that the relative difference between $v_2(p_T)$ in p+Au and d+Au collisions is smaller in our calculation than in the experimental data, but comparable given the experimental errors. Results for Au+Au collisions show a smaller $v_2$ than in d+Au collisions for most values of $p_T$.

We next compute the four-particle cumulants $c_2\{4\}$ that reduce non-flow contributions from intrinsic two- and three-particle correlations without the need for a large rapidity gap. In order to facilitate a comparison with the PHENIX data we consider momentum integrated quantities plotted against the charged particle multiplicity measured at forward rapidity, $N_{\rm tracks}^{\rm FVTX}$.\footnote{We use the following conversion relation to estimate forward multiplicity in our approach: $N^{\rm FVTX}_{\rm tracks}=1.96\, dN_{\rm ch}/d\eta$.}
Our results for all small systems are shown in Fig.\,\ref{fig:c24}. We find that the $c_2\{4\}$, that measures the momentum integrated anisotropy, is equal (within statistical errors) in p+Au and d+Au collisions for the same multiplicity. This shows that in our approach the difference observed between p+Au and d+Au for differential $v_2(p_T)$ seen in Fig.\,\ref{fig:v2pT} is compensated by the difference in $\langle p_T\rangle$ shown in Fig.\,\ref{fig:mpt}. Within errors, $c_2\{4\}$ in $^3$He+Au collisions also agrees with those in p+Au and d+Au collisions.

While in d+Au collisions $c_2\{4\}$ is negative both in the calculation (except for the lowest multiplicities) and the experimental data \cite{Aidala:2017pup}, in p+Au collisions it is positive in the experimental data, but negative in our calculation. The computed $c_2\{4\}$ in d+Au has a stronger multiplicity dependence than the experimental data, and is comparable to the AMPT result shown in \cite{Aidala:2017ajz}. The mean value of our result for 0-5\% central p+p collisions is positive, but within our statistics it is consistent with zero.

\paragraph{Role of geometry and initial momentum anisotropy.}
We compute the initial momentum anisotropy defined by
\begin{equation}\label{eq:epsilonp}
    \vec{\mathcal{E}}_p \equiv \varepsilon_p e^{i 2 \Psi_2^p} \equiv \frac{\langle T^{xx}-T^{yy}\rangle + i\langle 2 T^{xy}\rangle}{\langle T^{xx}+T^{yy}\rangle}\,,
\end{equation}
which is a proxy for the purely initial state $v_2$ (of the single-particle distribution) \cite{Krasnitz:2002ng}. Note that $\langle\cdot\rangle$ is defined without energy density weight.

\begin{figure*}[t]
  \includegraphics[width=1\textwidth]{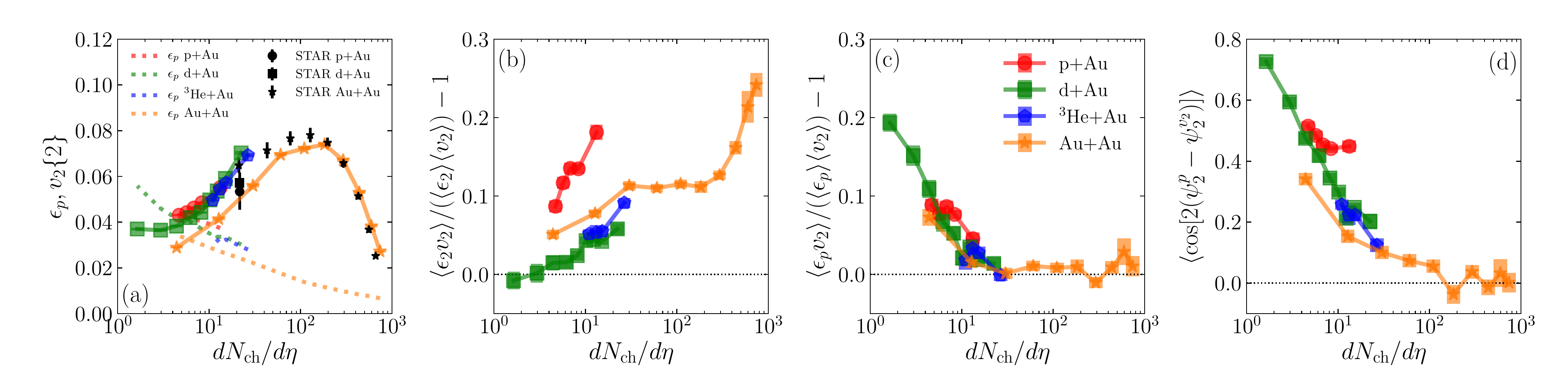}
  \caption{a) Initial momentum anisotropy and final $v_2$ as functions of multiplicity compared to STAR data \cite{Adam:2019woz}. STAR determined $dN_{\rm ch}/d\eta$ using a specific quark Monte-Carlo Glauber model. b) Correlation between the initial ellipticity and final $v_2$. c) Correlation between the magnitude of the initial momentum anisotropy $\varepsilon_p$ and final $v_2$. d) Correlation between the direction of the initial momentum anisotropy (given by $\psi_2^p$) and the final $v_2$ (with $\psi_2^{v_2}$ obtained from the flow vector). \label{fig:ep-v2-combined}}
\end{figure*}

We present results for $\varepsilon_p$ as a function of multiplicity for different systems together with final $v_2\{2\}$ values, which are compared to STAR data \cite{STAR:2019xzd,Adams:2004bi}, in Fig.\,\ref{fig:ep-v2-combined} (a).
The trend with multiplicity of the initial $\varepsilon_p$ is opposite to that of $v_2$.  This is in line with the color domain interpretation of the emergence of initial state anisotropies \cite{Dumitru:2014dra,Dumitru:2014vka,Dumitru:2014yza,Skokov:2014tka,Lappi:2015vta}, where the anisotropy is suppressed in larger systems. The effect of the hydrodynamic final state interactions reverses this trend, leading to increasing $v_2$ with increasing multiplicity.

The initial anisotropy $\varepsilon_p$ for p+Au collisions is larger than that for d+Au and Au+Au collisions at the same multiplicity, which is also obtained in other CGC calculations \cite{Mace:2018vwq} (for p+Au and d+Au) and is expected from the color domain interpretation. 
We note that the direct CGC calculation of two-particle correlations also includes contributions to the anisotropy from quantum interference effects, which are lost when taking $T^{\mu\nu}$ and inserting it into hydrodynamics. 

For Au+Au collisions we find smaller $v_2$ than for the "smaller" systems at the same multiplicity, in line with the result in Fig.\ref{fig:v2pT}. Given that $\varepsilon_p$ is larger in d+Au than in Au+Au collisions at the same multiplicity (and the eccentricities are very similar in the two systems, with $\varepsilon_2$ being slightly larger in Au+Au collisions), the fact that $v_2(p_T)$ is larger in d+Au collisions than in Au+Au collisions can be interpreted as an effect of the initial momentum anisotropy. We will show below that the final $v_2$ can in fact retain information on the initial momentum anisotropy for this multiplicity (Fig.\,\ref{fig:ep-v2-combined} (c) and (d)).

The STAR data \cite{Adam:2019woz} have large error bars, but the trend of the ordering between systems is opposite to our result. Because the STAR measurements were performed with a limited pseudorapidity separation of $|\Delta\eta|>0.7$, it is  expected that there are still non-flow contributions in the data. 
Because the difference between $v_2\{2\}$ in d+Au and Au+Au collisions at the same multiplicity could be a signal of the CGC momentum anisotropy, precise data on $v_n$ in d/p+Au and peripheral Au+Au collisions for the same kinematic cuts at RHIC are highly desirable. 

At LHC, the $v_2$ in $\sqrt{s}=5.02\,{\rm TeV}$ p+Pb collisions was found to be smaller than that in Pb+Pb collisions at the same multiplicity \cite{Chatrchyan:2013nka,Abelev:2014mda,Adam:2016izf}. This is not inconsistent with our results as at the higher LHC energy hydrodynamics gains in relative importance because higher energy densities lead to longer fireball lifetimes. We will explore this collision energy dependence quantitatively within our framework in the future.

Fig.\,\ref{fig:ep-v2-combined} (b) shows the correlation between the spatial ellipticity and the $v_2$, demonstrating the expected increasing trend with multiplicity. For p+Au collisions the correlation is stronger than for the other systems at the same multiplicity, probably because the measure of eccentricity works best as a predictor for $v_2$ for a system that is spatially connected - the other systems at low multiplicity often consist of separated regions that may not interact much with one another during the hydrodynamic evolution.

Fig.\,\ref{fig:ep-v2-combined} (c) shows the effect of the initial state momentum anisotropy on the observed $v_2$ by presenting the correlator of $\varepsilon_p$ and $v_2$ as a function of multiplicity. Results for different systems are very similar, all showing a significant correlation for systems with less than 10 charged hadrons per unit rapidity, with the correlation decreasing with increasing multiplicity, as the effects of hydrodynamic response to the initial spatial geometry become dominant. For Au+Au collisions more central than 50\%, the magnitude of the initial momentum anisotropy is uncorrelated with the final $v_2$.

The correlation between the direction of the initial state momentum anisotropy $\psi_2^p$ and the final elliptic flow is shown in Fig.\,\ref{fig:ep-v2-combined} (d). For small multiplicities the directions of the final flow and initial state momentum anisotropy are strongly correlated, with the correlation decreasing with increasing multiplicity as for the case of the magnitudes. This qualitatively agrees with the recent study using AMPT \cite{Nie:2019swk}. All small systems have a finite correlation, indicating that effects of the initial momentum anisotropy should never be neglected. For Au+Au collisions with $dN_{\rm ch}/d\eta\gtrsim 100$, the flow direction has no correlation with the initial momentum anisotropy anymore. Another interesting feature is that the correlation is stronger in p+Au than in d+Au collisions for the same multiplicity. This can be caused by a) the larger initial momentum anisotropy in p+Au and b) the differences in the hydrodynamic evolution, as system sizes are different in p+Au and d+Au at the same multiplicity. 

\begin{figure}[t]
\includegraphics[width=0.45\textwidth]{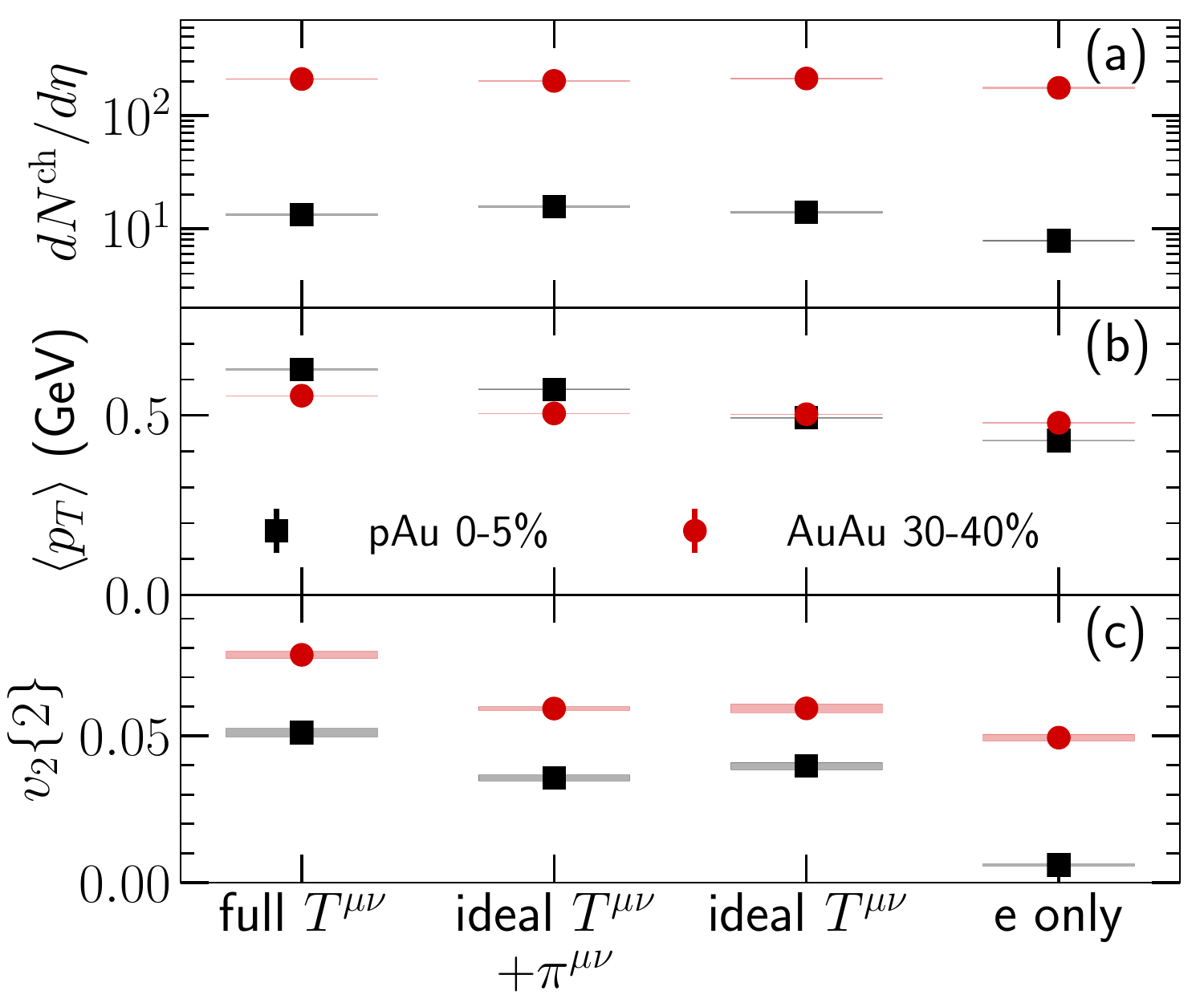}
  \caption{The effect of initial state features on observables. See text for details. \label{fig:pAu_initialsensitivity}}
\end{figure}

\paragraph{Effects of initial flow and viscous stress.}
To further explore the effect of the details in the initial state on observables, in Fig.\,\ref{fig:pAu_initialsensitivity} we show results for (a) $dN_{\rm ch}/d\eta$, (b) $\langle p_T \rangle$, and (c) $v_2\{2\}$ obtained with our standard calculation using the full energy momentum tensor and matching of the initial pressure to three other situations, where we i) exclude the initial $\Pi$ ii) exclude initial $\Pi$ and initial $\pi^{\mu\nu}$, and iii) exclude initial $\Pi$, initial $\pi^{\mu\nu}$ and initial flow $u^\mu$. We study both 30-40\% central Au+Au and 0-5\% central p+Au collisions. 

As expected, the effects of the changes to the initial state affect p+Au collisions more than the studied Au+Au collisions. The multiplicity varies by maximally 20\% in Au+Au collisions, mainly because removing the initial flow removes some of the deposited energy. In p+Au collisions, the maximal modification of $dN_{\rm ch}/d\eta$ is $\sim 40\%$. For Au+Au, $\langle p_T\rangle$ changes by maximally 15\%, in p+Au it can change by 30\%.\footnote{We note that given the choice of $(\zeta/s)(T)$ made in \cite{Schenke:2019ruo}, the bulk relaxation time ranges between 1 and $5\,{\rm fm}$ for temperatures between 200 and $400\,{\rm MeV}$. This can increase the large effect of the initial $\Pi$.} Other choices for $(\zeta/s)(T)$, e.g. with a narrower peak around $T_c$, would lead to a reduced $\tau_\Pi$ and a weakened effect of the initial $\Pi$.

The most significant changes we find for $v_2$, which can change by $\sim 35\%$ in Au+Au collisions and by as much as $90\%$ in p+Au. This enormous effect in p+Au is a result of very little flow build-up in the absence of initial flow and presence of a large bulk viscosity. While none of the scenarios, other than using the full $T^{\mu\nu}$, are fully physical, because they do not conserve energy and momentum at the time of switching to hydrodynamics, they give us a good estimate on how much the details of the initial state, other than the geometry, matter in small and larger systems. Our results suggest that for any initial state model, inclusion of initial flow or off-equilibrium contributions will affect final observables in small systems. Merely using spatial distributions of energy densities is likely too crude an approximation, for some observables even in Au+Au collisions. 

An uncertainty we have not yet discussed stems from the choice of $\tau_{\rm init}$. We found that in typical (30-40\%) Au+Au collisions $\langle p_T \rangle$, $v_2\{2\}$, and $v_3\{2\}$ all change by less than 1\% when changing $\tau_{\rm init}$ to 0.2\,{\rm fm}. The biggest change was an increase of $dN_{\rm ch}/d\eta$ by 20\%, caused by increased viscous entropy production. On the other hand, 0-5\% p+Au collisions show more sensitivity to $\tau_{\rm init}$. Most observables changed between 3 and 15\% from the results at $\tau_{\rm init}=0.4\,{\rm fm}$ varying it up or down by $0.2\,{\rm fm}$. The two larger changes were an increase of $dN_{\rm ch}/d\eta$ by 35\% and a decrease of $v_3\{2\}$ by 22\% when going to  $\tau_{\rm init}=0.2\,{\rm fm}$.

\paragraph{Conclusions.} We have presented calculations for multi-particle correlation observables in systems studied in the RHIC small system scan within a hybrid framework including an initial state from the Color Glass Condensate effective theory, hydrodynamic evolution, and microscopic hadronic transport. All free parameters of the model were constrained by Au+Au collisions previously \cite{Schenke:2019ruo}. Our framework includes CGC initial state momentum anisotropies and final state response to the initial spatial geometry. We demonstrated that the two mechanisms lead to characteristically different multiplicity dependencies of the momentum anisotropy and that final state effects are necessary to describe the qualitative features of the experimentally observed anisotropies. 

This work is the first to have quantified the amount that the initial state momentum anisotropy can contribute to observables, if final state interactions are described by realistic hydrodynamic simulations. We have found non-zero correlations between the initial momentum anisotropy magnitude and the charged hadron $v_2$, as well as their orientations. The correlation decreases with increasing multiplicity and vanishes in large systems (Au+Au, more central than $\sim 50\%$). Generally, the details of the initial state, including the initial flow profile and viscous stress tensor, affect the quantitative results for experimental observables strongly.

We overestimate experimental results for $v_2(p_T)$ and $v_3(p_T)$ from PHENIX, which could have to do with the fact that our calculations are boost-invariant, and consequently cannot employ exactly the same method as the experiment. It is also possible that a better set of parameters could be found that yields good agreement with both heavy ion and small system collision data. 

While the $p_T$ dependent $v_2$ in p+Au and d+Au collisions show the expected ordering from geometry at the same multiplicity, the fact that $v_2\{2\}$ in d+Au has a larger value than Au+Au at the same multiplicity is interpreted as a signature of the initial state momentum anisotropy. It does not result from differences in $\langle p_T \rangle$, because $v_2(p_T)$ shows the same trend, and it is not a consequence of different eccentricities, which are close and ordered in the opposite way. Consequently, precise measurements of $v_2$ for these two systems at RHIC energy and at equal multiplicities would be highly desirable to help unveil a possible signature of the initial state momentum anisotropy.

Further progress will require 3D simulations either using extensions of the IP-Glasma initial state \cite{Schenke:2016ksl,McDonald:2018wql} or alternative three dimensional initial state models \cite{Shen:2017bsr}, which can also be applied to lower energy collisions, like the ones in the d+Au energy scan \cite{Aidala:2017pup}, but lack the initial state momentum anisotropy emphasized here. Better constraints on QCD transport coefficients, improved descriptions of the intermediate regime between Yang-Mills and hydrodynamic stages (see e.g. \cite{Kurkela:2018wud, Kurkela:2018vqr}), as well as improved particlization mechanisms (see e.g. \cite{Oliinychenko:2019zfk}), and inclusion of hydrodynamic fluctuations \cite{Singh:2018dpk} are potentially important future steps for the description of small system collisions.

{\bf Acknowledgments} 
We thank Ron Belmont for his help converting to PHENIX's multiplicity measure. We thank Heikki M\"antysaari and Raju Venugopalan for helpful discussions.
B.P.S. and P.T. are supported under DOE Contract No. DE-SC0012704. C.S. is supported under DOE Contract No. DE-SC0013460. This research used resources of the National Energy Research Scientific Computing Center, which is supported by the Office of Science of the U.S. Department of Energy under Contract No. DE-AC02-05CH11231 and resources of the high performance computing services at Wayne State University. This work is supported in part by the U.S. Department of Energy, Office of Science, Office of Nuclear Physics, within the framework of the Beam Energy Scan Theory (BEST) Topical Collaboration.

\bibliography{spires}

\begin{thebibliography}{91}%
\makeatletter
\providecommand \@ifxundefined [1]{%
 \@ifx{#1\undefined}
}%
\providecommand \@ifnum [1]{%
 \ifnum #1\expandafter \@firstoftwo
 \else \expandafter \@secondoftwo
 \fi
}%
\providecommand \@ifx [1]{%
 \ifx #1\expandafter \@firstoftwo
 \else \expandafter \@secondoftwo
 \fi
}%
\providecommand \natexlab [1]{#1}%
\providecommand \enquote  [1]{``#1''}%
\providecommand \bibnamefont  [1]{#1}%
\providecommand \bibfnamefont [1]{#1}%
\providecommand \citenamefont [1]{#1}%
\providecommand \href@noop [0]{\@secondoftwo}%
\providecommand \href [0]{\begingroup \@sanitize@url \@href}%
\providecommand \@href[1]{\@@startlink{#1}\@@href}%
\providecommand \@@href[1]{\endgroup#1\@@endlink}%
\providecommand \@sanitize@url [0]{\catcode `\\12\catcode `\$12\catcode
  `\&12\catcode `\#12\catcode `\^12\catcode `\_12\catcode `\%12\relax}%
\providecommand \@@startlink[1]{}%
\providecommand \@@endlink[0]{}%
\providecommand \url  [0]{\begingroup\@sanitize@url \@url }%
\providecommand \@url [1]{\endgroup\@href {#1}{\urlprefix }}%
\providecommand \urlprefix  [0]{URL }%
\providecommand \Eprint [0]{\href }%
\providecommand \doibase [0]{http://dx.doi.org/}%
\providecommand \selectlanguage [0]{\@gobble}%
\providecommand \bibinfo  [0]{\@secondoftwo}%
\providecommand \bibfield  [0]{\@secondoftwo}%
\providecommand \translation [1]{[#1]}%
\providecommand \BibitemOpen [0]{}%
\providecommand \bibitemStop [0]{}%
\providecommand \bibitemNoStop [0]{.\EOS\space}%
\providecommand \EOS [0]{\spacefactor3000\relax}%
\providecommand \BibitemShut  [1]{\csname bibitem#1\endcsname}%
\let\auto@bib@innerbib\@empty
\bibitem [{\citenamefont {Dusling}\ \emph {et~al.}(2016)\citenamefont
  {Dusling}, \citenamefont {Li},\ and\ \citenamefont
  {Schenke}}]{Dusling:2015gta}%
  \BibitemOpen
  \bibfield  {author} {\bibinfo {author} {\bibfnamefont {K.}~\bibnamefont
  {Dusling}}, \bibinfo {author} {\bibfnamefont {W.}~\bibnamefont {Li}}, \ and\
  \bibinfo {author} {\bibfnamefont {B.}~\bibnamefont {Schenke}},\ }\href
  {\doibase 10.1142/S0218301316300022} {\bibfield  {journal} {\bibinfo
  {journal} {Int. J. Mod. Phys.}\ }\textbf {\bibinfo {volume} {E25}},\ \bibinfo
  {pages} {1630002} (\bibinfo {year} {2016})},\ \Eprint
  {http://arxiv.org/abs/1509.07939} {arXiv:1509.07939 [nucl-ex]} \BibitemShut
  {NoStop}%
\bibitem [{\citenamefont {Schlichting}\ and\ \citenamefont
  {Tribedy}(2016)}]{Schlichting:2016kjw}%
  \BibitemOpen
  \bibfield  {author} {\bibinfo {author} {\bibfnamefont {S.}~\bibnamefont
  {Schlichting}}\ and\ \bibinfo {author} {\bibfnamefont {P.}~\bibnamefont
  {Tribedy}},\ }\href {\doibase 10.1155/2016/8460349} {\bibfield  {journal}
  {\bibinfo  {journal} {Adv. High Energy Phys.}\ }\textbf {\bibinfo {volume}
  {2016}},\ \bibinfo {pages} {8460349} (\bibinfo {year} {2016})}\BibitemShut
  {NoStop}%
\bibitem [{\citenamefont {Dumitru}\ \emph {et~al.}(2008)\citenamefont
  {Dumitru}, \citenamefont {Gelis}, \citenamefont {McLerran},\ and\
  \citenamefont {Venugopalan}}]{Dumitru:2008wn}%
  \BibitemOpen
  \bibfield  {author} {\bibinfo {author} {\bibfnamefont {A.}~\bibnamefont
  {Dumitru}}, \bibinfo {author} {\bibfnamefont {F.}~\bibnamefont {Gelis}},
  \bibinfo {author} {\bibfnamefont {L.}~\bibnamefont {McLerran}}, \ and\
  \bibinfo {author} {\bibfnamefont {R.}~\bibnamefont {Venugopalan}},\
  }\href@noop {} {\  (\bibinfo {year} {2008})},\ \Eprint
  {http://arxiv.org/abs/0804.3858} {arXiv:0804.3858 [hep-ph]} \BibitemShut
  {NoStop}%
\bibitem [{\citenamefont {Kovner}\ and\ \citenamefont
  {Lublinsky}(2011{\natexlab{a}})}]{Kovner:2010xk}%
  \BibitemOpen
  \bibfield  {author} {\bibinfo {author} {\bibfnamefont {A.}~\bibnamefont
  {Kovner}}\ and\ \bibinfo {author} {\bibfnamefont {M.}~\bibnamefont
  {Lublinsky}},\ }\href {\doibase 10.1103/PhysRevD.83.034017} {\bibfield
  {journal} {\bibinfo  {journal} {Phys. Rev.}\ }\textbf {\bibinfo {volume}
  {D83}},\ \bibinfo {pages} {034017} (\bibinfo {year} {2011}{\natexlab{a}})},\
  \Eprint {http://arxiv.org/abs/1012.3398} {arXiv:1012.3398 [hep-ph]}
  \BibitemShut {NoStop}%
\bibitem [{\citenamefont {Dumitru}\ \emph {et~al.}(2011)\citenamefont
  {Dumitru}, \citenamefont {Dusling}, \citenamefont {Gelis}, \citenamefont
  {Jalilian-Marian}, \citenamefont {Lappi},\ and\ \citenamefont
  {Venugopalan}}]{Dumitru:2010iy}%
  \BibitemOpen
  \bibfield  {author} {\bibinfo {author} {\bibfnamefont {A.}~\bibnamefont
  {Dumitru}}, \bibinfo {author} {\bibfnamefont {K.}~\bibnamefont {Dusling}},
  \bibinfo {author} {\bibfnamefont {F.}~\bibnamefont {Gelis}}, \bibinfo
  {author} {\bibfnamefont {J.}~\bibnamefont {Jalilian-Marian}}, \bibinfo
  {author} {\bibfnamefont {T.}~\bibnamefont {Lappi}}, \ and\ \bibinfo {author}
  {\bibfnamefont {R.}~\bibnamefont {Venugopalan}},\ }\href {\doibase
  10.1016/j.physletb.2011.01.024} {\bibfield  {journal} {\bibinfo  {journal}
  {Phys. Lett.}\ }\textbf {\bibinfo {volume} {B697}},\ \bibinfo {pages} {21}
  (\bibinfo {year} {2011})},\ \Eprint {http://arxiv.org/abs/1009.5295}
  {arXiv:1009.5295 [hep-ph]} \BibitemShut {NoStop}%
\bibitem [{\citenamefont {Kovner}\ and\ \citenamefont
  {Lublinsky}(2011{\natexlab{b}})}]{Kovner:2011pe}%
  \BibitemOpen
  \bibfield  {author} {\bibinfo {author} {\bibfnamefont {A.}~\bibnamefont
  {Kovner}}\ and\ \bibinfo {author} {\bibfnamefont {M.}~\bibnamefont
  {Lublinsky}},\ }\href {\doibase 10.1103/PhysRevD.84.094011} {\bibfield
  {journal} {\bibinfo  {journal} {Phys. Rev.}\ }\textbf {\bibinfo {volume}
  {D84}},\ \bibinfo {pages} {094011} (\bibinfo {year} {2011}{\natexlab{b}})},\
  \Eprint {http://arxiv.org/abs/1109.0347} {arXiv:1109.0347 [hep-ph]}
  \BibitemShut {NoStop}%
\bibitem [{\citenamefont {Dusling}\ and\ \citenamefont
  {Venugopalan}(2012)}]{Dusling:2012iga}%
  \BibitemOpen
  \bibfield  {author} {\bibinfo {author} {\bibfnamefont {K.}~\bibnamefont
  {Dusling}}\ and\ \bibinfo {author} {\bibfnamefont {R.}~\bibnamefont
  {Venugopalan}},\ }\href {\doibase 10.1103/PhysRevLett.108.262001} {\bibfield
  {journal} {\bibinfo  {journal} {Phys.Rev.Lett.}\ }\textbf {\bibinfo {volume}
  {108}},\ \bibinfo {pages} {262001} (\bibinfo {year} {2012})},\ \Eprint
  {http://arxiv.org/abs/1201.2658} {arXiv:1201.2658 [hep-ph]} \BibitemShut
  {NoStop}%
\bibitem [{\citenamefont {Levin}\ and\ \citenamefont
  {Rezaeian}(2011)}]{Levin:2011fb}%
  \BibitemOpen
  \bibfield  {author} {\bibinfo {author} {\bibfnamefont {E.}~\bibnamefont
  {Levin}}\ and\ \bibinfo {author} {\bibfnamefont {A.~H.}\ \bibnamefont
  {Rezaeian}},\ }\href {\doibase 10.1103/PhysRevD.84.034031} {\bibfield
  {journal} {\bibinfo  {journal} {Phys. Rev.}\ }\textbf {\bibinfo {volume}
  {D84}},\ \bibinfo {pages} {034031} (\bibinfo {year} {2011})},\ \Eprint
  {http://arxiv.org/abs/1105.3275} {arXiv:1105.3275 [hep-ph]} \BibitemShut
  {NoStop}%
\bibitem [{\citenamefont {Dusling}\ and\ \citenamefont
  {Venugopalan}(2013{\natexlab{a}})}]{Dusling:2012wy}%
  \BibitemOpen
  \bibfield  {author} {\bibinfo {author} {\bibfnamefont {K.}~\bibnamefont
  {Dusling}}\ and\ \bibinfo {author} {\bibfnamefont {R.}~\bibnamefont
  {Venugopalan}},\ }\href {\doibase 10.1103/PhysRevD.87.054014} {\bibfield
  {journal} {\bibinfo  {journal} {Phys. Rev.}\ }\textbf {\bibinfo {volume}
  {D87}},\ \bibinfo {pages} {054014} (\bibinfo {year} {2013}{\natexlab{a}})},\
  \Eprint {http://arxiv.org/abs/1211.3701} {arXiv:1211.3701 [hep-ph]}
  \BibitemShut {NoStop}%
\bibitem [{\citenamefont {Dusling}\ and\ \citenamefont
  {Venugopalan}(2013{\natexlab{b}})}]{Dusling:2013qoz}%
  \BibitemOpen
  \bibfield  {author} {\bibinfo {author} {\bibfnamefont {K.}~\bibnamefont
  {Dusling}}\ and\ \bibinfo {author} {\bibfnamefont {R.}~\bibnamefont
  {Venugopalan}},\ }\href {\doibase 10.1103/PhysRevD.87.094034} {\bibfield
  {journal} {\bibinfo  {journal} {Phys. Rev.}\ }\textbf {\bibinfo {volume}
  {D87}},\ \bibinfo {pages} {094034} (\bibinfo {year} {2013}{\natexlab{b}})},\
  \Eprint {http://arxiv.org/abs/1302.7018} {arXiv:1302.7018 [hep-ph]}
  \BibitemShut {NoStop}%
\bibitem [{\citenamefont {Dumitru}\ and\ \citenamefont
  {Giannini}(2015)}]{Dumitru:2014dra}%
  \BibitemOpen
  \bibfield  {author} {\bibinfo {author} {\bibfnamefont {A.}~\bibnamefont
  {Dumitru}}\ and\ \bibinfo {author} {\bibfnamefont {A.~V.}\ \bibnamefont
  {Giannini}},\ }\href {\doibase 10.1016/j.nuclphysa.2014.10.037} {\bibfield
  {journal} {\bibinfo  {journal} {Nucl. Phys.}\ }\textbf {\bibinfo {volume}
  {A933}},\ \bibinfo {pages} {212} (\bibinfo {year} {2015})},\ \Eprint
  {http://arxiv.org/abs/1406.5781} {arXiv:1406.5781 [hep-ph]} \BibitemShut
  {NoStop}%
\bibitem [{\citenamefont {Dumitru}\ \emph {et~al.}(2015)\citenamefont
  {Dumitru}, \citenamefont {McLerran},\ and\ \citenamefont
  {Skokov}}]{Dumitru:2014yza}%
  \BibitemOpen
  \bibfield  {author} {\bibinfo {author} {\bibfnamefont {A.}~\bibnamefont
  {Dumitru}}, \bibinfo {author} {\bibfnamefont {L.}~\bibnamefont {McLerran}}, \
  and\ \bibinfo {author} {\bibfnamefont {V.}~\bibnamefont {Skokov}},\ }\href
  {\doibase 10.1016/j.physletb.2015.02.046} {\bibfield  {journal} {\bibinfo
  {journal} {Phys. Lett.}\ }\textbf {\bibinfo {volume} {B743}},\ \bibinfo
  {pages} {134} (\bibinfo {year} {2015})},\ \Eprint
  {http://arxiv.org/abs/1410.4844} {arXiv:1410.4844 [hep-ph]} \BibitemShut
  {NoStop}%
\bibitem [{\citenamefont {Schenke}\ \emph {et~al.}(2015)\citenamefont
  {Schenke}, \citenamefont {Schlichting},\ and\ \citenamefont
  {Venugopalan}}]{Schenke:2015aqa}%
  \BibitemOpen
  \bibfield  {author} {\bibinfo {author} {\bibfnamefont {B.}~\bibnamefont
  {Schenke}}, \bibinfo {author} {\bibfnamefont {S.}~\bibnamefont
  {Schlichting}}, \ and\ \bibinfo {author} {\bibfnamefont {R.}~\bibnamefont
  {Venugopalan}},\ }\href {\doibase 10.1016/j.physletb.2015.05.051} {\bibfield
  {journal} {\bibinfo  {journal} {Phys. Lett.}\ }\textbf {\bibinfo {volume}
  {B747}},\ \bibinfo {pages} {76} (\bibinfo {year} {2015})},\ \Eprint
  {http://arxiv.org/abs/1502.01331} {arXiv:1502.01331 [hep-ph]} \BibitemShut
  {NoStop}%
\bibitem [{\citenamefont {McLerran}\ and\ \citenamefont
  {Skokov}(2016)}]{McLerran:2015sva}%
  \BibitemOpen
  \bibfield  {author} {\bibinfo {author} {\bibfnamefont {L.}~\bibnamefont
  {McLerran}}\ and\ \bibinfo {author} {\bibfnamefont {V.}~\bibnamefont
  {Skokov}},\ }\href {\doibase 10.1016/j.nuclphysa.2015.12.005} {\bibfield
  {journal} {\bibinfo  {journal} {Nucl. Phys.}\ }\textbf {\bibinfo {volume}
  {A947}},\ \bibinfo {pages} {142} (\bibinfo {year} {2016})},\ \Eprint
  {http://arxiv.org/abs/1510.08072} {arXiv:1510.08072 [hep-ph]} \BibitemShut
  {NoStop}%
\bibitem [{\citenamefont {Schenke}\ \emph {et~al.}(2016)\citenamefont
  {Schenke}, \citenamefont {Schlichting}, \citenamefont {Tribedy},\ and\
  \citenamefont {Venugopalan}}]{Schenke:2016lrs}%
  \BibitemOpen
  \bibfield  {author} {\bibinfo {author} {\bibfnamefont {B.}~\bibnamefont
  {Schenke}}, \bibinfo {author} {\bibfnamefont {S.}~\bibnamefont
  {Schlichting}}, \bibinfo {author} {\bibfnamefont {P.}~\bibnamefont
  {Tribedy}}, \ and\ \bibinfo {author} {\bibfnamefont {R.}~\bibnamefont
  {Venugopalan}},\ }\href {\doibase 10.1103/PhysRevLett.117.162301} {\bibfield
  {journal} {\bibinfo  {journal} {Phys. Rev. Lett.}\ }\textbf {\bibinfo
  {volume} {117}},\ \bibinfo {pages} {162301} (\bibinfo {year} {2016})},\
  \Eprint {http://arxiv.org/abs/1607.02496} {arXiv:1607.02496 [hep-ph]}
  \BibitemShut {NoStop}%
\bibitem [{\citenamefont {Dusling}\ \emph
  {et~al.}(2018{\natexlab{a}})\citenamefont {Dusling}, \citenamefont {Mace},\
  and\ \citenamefont {Venugopalan}}]{Dusling:2017dqg}%
  \BibitemOpen
  \bibfield  {author} {\bibinfo {author} {\bibfnamefont {K.}~\bibnamefont
  {Dusling}}, \bibinfo {author} {\bibfnamefont {M.}~\bibnamefont {Mace}}, \
  and\ \bibinfo {author} {\bibfnamefont {R.}~\bibnamefont {Venugopalan}},\
  }\href {\doibase 10.1103/PhysRevLett.120.042002} {\bibfield  {journal}
  {\bibinfo  {journal} {Phys. Rev. Lett.}\ }\textbf {\bibinfo {volume} {120}},\
  \bibinfo {pages} {042002} (\bibinfo {year} {2018}{\natexlab{a}})},\ \Eprint
  {http://arxiv.org/abs/1705.00745} {arXiv:1705.00745 [hep-ph]} \BibitemShut
  {NoStop}%
\bibitem [{\citenamefont {Dusling}\ \emph
  {et~al.}(2018{\natexlab{b}})\citenamefont {Dusling}, \citenamefont {Mace},\
  and\ \citenamefont {Venugopalan}}]{Dusling:2017aot}%
  \BibitemOpen
  \bibfield  {author} {\bibinfo {author} {\bibfnamefont {K.}~\bibnamefont
  {Dusling}}, \bibinfo {author} {\bibfnamefont {M.}~\bibnamefont {Mace}}, \
  and\ \bibinfo {author} {\bibfnamefont {R.}~\bibnamefont {Venugopalan}},\
  }\href {\doibase 10.1103/PhysRevD.97.016014} {\bibfield  {journal} {\bibinfo
  {journal} {Phys. Rev.}\ }\textbf {\bibinfo {volume} {D97}},\ \bibinfo {pages}
  {016014} (\bibinfo {year} {2018}{\natexlab{b}})},\ \Eprint
  {http://arxiv.org/abs/1706.06260} {arXiv:1706.06260 [hep-ph]} \BibitemShut
  {NoStop}%
\bibitem [{\citenamefont {Mace}\ \emph {et~al.}(2018)\citenamefont {Mace},
  \citenamefont {Skokov}, \citenamefont {Tribedy},\ and\ \citenamefont
  {Venugopalan}}]{Mace:2018vwq}%
  \BibitemOpen
  \bibfield  {author} {\bibinfo {author} {\bibfnamefont {M.}~\bibnamefont
  {Mace}}, \bibinfo {author} {\bibfnamefont {V.~V.}\ \bibnamefont {Skokov}},
  \bibinfo {author} {\bibfnamefont {P.}~\bibnamefont {Tribedy}}, \ and\
  \bibinfo {author} {\bibfnamefont {R.}~\bibnamefont {Venugopalan}},\ }\href
  {\doibase 10.1103/PhysRevLett.123.039901, 10.1103/PhysRevLett.121.052301}
  {\bibfield  {journal} {\bibinfo  {journal} {Phys. Rev. Lett.}\ }\textbf
  {\bibinfo {volume} {121}},\ \bibinfo {pages} {052301} (\bibinfo {year}
  {2018})},\ \bibinfo {note} {[Erratum: Phys. Rev.
  Lett.123,no.3,039901(2019)]},\ \Eprint {http://arxiv.org/abs/1805.09342}
  {arXiv:1805.09342 [hep-ph]} \BibitemShut {NoStop}%
\bibitem [{\citenamefont {Mace}\ \emph {et~al.}(2019)\citenamefont {Mace},
  \citenamefont {Skokov}, \citenamefont {Tribedy},\ and\ \citenamefont
  {Venugopalan}}]{Mace:2018yvl}%
  \BibitemOpen
  \bibfield  {author} {\bibinfo {author} {\bibfnamefont {M.}~\bibnamefont
  {Mace}}, \bibinfo {author} {\bibfnamefont {V.~V.}\ \bibnamefont {Skokov}},
  \bibinfo {author} {\bibfnamefont {P.}~\bibnamefont {Tribedy}}, \ and\
  \bibinfo {author} {\bibfnamefont {R.}~\bibnamefont {Venugopalan}},\ }\href
  {\doibase 10.1016/j.physletb.2018.09.064} {\bibfield  {journal} {\bibinfo
  {journal} {Phys. Lett.}\ }\textbf {\bibinfo {volume} {B788}},\ \bibinfo
  {pages} {161} (\bibinfo {year} {2019})},\ \Eprint
  {http://arxiv.org/abs/1807.00825} {arXiv:1807.00825 [hep-ph]} \BibitemShut
  {NoStop}%
\bibitem [{\citenamefont {Kovner}\ and\ \citenamefont
  {Skokov}(2018)}]{Kovner:2018fxj}%
  \BibitemOpen
  \bibfield  {author} {\bibinfo {author} {\bibfnamefont {A.}~\bibnamefont
  {Kovner}}\ and\ \bibinfo {author} {\bibfnamefont {V.~V.}\ \bibnamefont
  {Skokov}},\ }\href {\doibase 10.1016/j.physletb.2018.09.001} {\bibfield
  {journal} {\bibinfo  {journal} {Phys. Lett.}\ }\textbf {\bibinfo {volume}
  {B785}},\ \bibinfo {pages} {372} (\bibinfo {year} {2018})},\ \Eprint
  {http://arxiv.org/abs/1805.09297} {arXiv:1805.09297 [hep-ph]} \BibitemShut
  {NoStop}%
\bibitem [{\citenamefont {McLerran}\ and\ \citenamefont
  {Venugopalan}(1994{\natexlab{a}})}]{McLerran:1994ni}%
  \BibitemOpen
  \bibfield  {author} {\bibinfo {author} {\bibfnamefont {L.~D.}\ \bibnamefont
  {McLerran}}\ and\ \bibinfo {author} {\bibfnamefont {R.}~\bibnamefont
  {Venugopalan}},\ }\href {\doibase 10.1103/PhysRevD.49.2233} {\bibfield
  {journal} {\bibinfo  {journal} {Phys. Rev.}\ }\textbf {\bibinfo {volume}
  {D49}},\ \bibinfo {pages} {2233} (\bibinfo {year}
  {1994}{\natexlab{a}})}\BibitemShut {NoStop}%
\bibitem [{\citenamefont {McLerran}\ and\ \citenamefont
  {Venugopalan}(1994{\natexlab{b}})}]{McLerran:1994ka}%
  \BibitemOpen
  \bibfield  {author} {\bibinfo {author} {\bibfnamefont {L.~D.}\ \bibnamefont
  {McLerran}}\ and\ \bibinfo {author} {\bibfnamefont {R.}~\bibnamefont
  {Venugopalan}},\ }\href {\doibase 10.1103/PhysRevD.49.3352} {\bibfield
  {journal} {\bibinfo  {journal} {Phys. Rev.}\ }\textbf {\bibinfo {volume}
  {D49}},\ \bibinfo {pages} {3352} (\bibinfo {year}
  {1994}{\natexlab{b}})}\BibitemShut {NoStop}%
\bibitem [{\citenamefont {Iancu}\ and\ \citenamefont
  {Venugopalan}(2003)}]{Iancu:2003xm}%
  \BibitemOpen
  \bibfield  {author} {\bibinfo {author} {\bibfnamefont {E.}~\bibnamefont
  {Iancu}}\ and\ \bibinfo {author} {\bibfnamefont {R.}~\bibnamefont
  {Venugopalan}},\ }in\ \href@noop {} {\emph {\bibinfo {booktitle} {Quark gluon
  plasma}}},\ \bibinfo {editor} {edited by\ \bibinfo {editor} {\bibfnamefont
  {R.}~\bibnamefont {Hwa}}\ and\ \bibinfo {editor} {\bibfnamefont {X.~N.}\
  \bibnamefont {Wang}}}\ (\bibinfo  {publisher} {World Scientific},\ \bibinfo
  {year} {2003})\ \Eprint {http://arxiv.org/abs/hep-ph/0303204}
  {hep-ph/0303204} \BibitemShut {NoStop}%
\bibitem [{\citenamefont {Bozek}(2012)}]{Bozek:2011if}%
  \BibitemOpen
  \bibfield  {author} {\bibinfo {author} {\bibfnamefont {P.}~\bibnamefont
  {Bozek}},\ }\href {\doibase 10.1103/PhysRevC.85.014911} {\bibfield  {journal}
  {\bibinfo  {journal} {Phys.Rev.}\ }\textbf {\bibinfo {volume} {C85}},\
  \bibinfo {pages} {014911} (\bibinfo {year} {2012})},\ \Eprint
  {http://arxiv.org/abs/1112.0915} {arXiv:1112.0915 [hep-ph]} \BibitemShut
  {NoStop}%
\bibitem [{\citenamefont {Bozek}\ and\ \citenamefont
  {Broniowski}(2013{\natexlab{a}})}]{Bozek:2012gr}%
  \BibitemOpen
  \bibfield  {author} {\bibinfo {author} {\bibfnamefont {P.}~\bibnamefont
  {Bozek}}\ and\ \bibinfo {author} {\bibfnamefont {W.}~\bibnamefont
  {Broniowski}},\ }\href {\doibase 10.1016/j.physletb.2012.12.051} {\bibfield
  {journal} {\bibinfo  {journal} {Phys.Lett.}\ }\textbf {\bibinfo {volume}
  {B718}},\ \bibinfo {pages} {1557} (\bibinfo {year} {2013}{\natexlab{a}})},\
  \Eprint {http://arxiv.org/abs/1211.0845} {arXiv:1211.0845 [nucl-th]}
  \BibitemShut {NoStop}%
\bibitem [{\citenamefont {Bozek}\ and\ \citenamefont
  {Broniowski}(2013{\natexlab{b}})}]{Bozek:2013df}%
  \BibitemOpen
  \bibfield  {author} {\bibinfo {author} {\bibfnamefont {P.}~\bibnamefont
  {Bozek}}\ and\ \bibinfo {author} {\bibfnamefont {W.}~\bibnamefont
  {Broniowski}},\ }\href@noop {} {\  (\bibinfo {year} {2013}{\natexlab{b}})},\
  \Eprint {http://arxiv.org/abs/1301.3314} {arXiv:1301.3314 [nucl-th]}
  \BibitemShut {NoStop}%
\bibitem [{\citenamefont {Bozek}\ and\ \citenamefont
  {Broniowski}(2013{\natexlab{c}})}]{Bozek:2013uha}%
  \BibitemOpen
  \bibfield  {author} {\bibinfo {author} {\bibfnamefont {P.}~\bibnamefont
  {Bozek}}\ and\ \bibinfo {author} {\bibfnamefont {W.}~\bibnamefont
  {Broniowski}},\ }\href {\doibase 10.1103/PhysRevC.88.014903} {\bibfield
  {journal} {\bibinfo  {journal} {Phys.Rev.}\ }\textbf {\bibinfo {volume}
  {C88}},\ \bibinfo {pages} {014903} (\bibinfo {year} {2013}{\natexlab{c}})},\
  \Eprint {http://arxiv.org/abs/1304.3044} {arXiv:1304.3044 [nucl-th]}
  \BibitemShut {NoStop}%
\bibitem [{\citenamefont {Bozek}\ \emph {et~al.}(2013)\citenamefont {Bozek},
  \citenamefont {Broniowski},\ and\ \citenamefont {Torrieri}}]{Bozek:2013ska}%
  \BibitemOpen
  \bibfield  {author} {\bibinfo {author} {\bibfnamefont {P.}~\bibnamefont
  {Bozek}}, \bibinfo {author} {\bibfnamefont {W.}~\bibnamefont {Broniowski}}, \
  and\ \bibinfo {author} {\bibfnamefont {G.}~\bibnamefont {Torrieri}},\ }\href
  {\doibase 10.1103/PhysRevLett.111.172303} {\bibfield  {journal} {\bibinfo
  {journal} {Phys.Rev.Lett.}\ }\textbf {\bibinfo {volume} {111}},\ \bibinfo
  {pages} {172303} (\bibinfo {year} {2013})}\BibitemShut {NoStop}%
\bibitem [{\citenamefont {Bzdak}\ \emph {et~al.}(2013)\citenamefont {Bzdak},
  \citenamefont {Schenke}, \citenamefont {Tribedy},\ and\ \citenamefont
  {Venugopalan}}]{Bzdak:2013zma}%
  \BibitemOpen
  \bibfield  {author} {\bibinfo {author} {\bibfnamefont {A.}~\bibnamefont
  {Bzdak}}, \bibinfo {author} {\bibfnamefont {B.}~\bibnamefont {Schenke}},
  \bibinfo {author} {\bibfnamefont {P.}~\bibnamefont {Tribedy}}, \ and\
  \bibinfo {author} {\bibfnamefont {R.}~\bibnamefont {Venugopalan}},\
  }\href@noop {} {\  (\bibinfo {year} {2013})},\ \Eprint
  {http://arxiv.org/abs/1304.3403} {arXiv:1304.3403 [nucl-th]} \BibitemShut
  {NoStop}%
\bibitem [{\citenamefont {Qin}\ and\ \citenamefont
  {Muller}(2014)}]{Qin:2013bha}%
  \BibitemOpen
  \bibfield  {author} {\bibinfo {author} {\bibfnamefont {G.-Y.}\ \bibnamefont
  {Qin}}\ and\ \bibinfo {author} {\bibfnamefont {B.}~\bibnamefont {Muller}},\
  }\href {\doibase 10.1103/PhysRevC.89.044902} {\bibfield  {journal} {\bibinfo
  {journal} {Phys.Rev.}\ }\textbf {\bibinfo {volume} {C89}},\ \bibinfo {pages}
  {044902} (\bibinfo {year} {2014})},\ \Eprint {http://arxiv.org/abs/1306.3439}
  {arXiv:1306.3439 [nucl-th]} \BibitemShut {NoStop}%
\bibitem [{\citenamefont {Werner}\ \emph {et~al.}(2014)\citenamefont {Werner},
  \citenamefont {Bleicher}, \citenamefont {Guiot}, \citenamefont {Karpenko},\
  and\ \citenamefont {Pierog}}]{Werner:2013ipa}%
  \BibitemOpen
  \bibfield  {author} {\bibinfo {author} {\bibfnamefont {K.}~\bibnamefont
  {Werner}}, \bibinfo {author} {\bibfnamefont {M.}~\bibnamefont {Bleicher}},
  \bibinfo {author} {\bibfnamefont {B.}~\bibnamefont {Guiot}}, \bibinfo
  {author} {\bibfnamefont {I.}~\bibnamefont {Karpenko}}, \ and\ \bibinfo
  {author} {\bibfnamefont {T.}~\bibnamefont {Pierog}},\ }\href {\doibase
  10.1103/PhysRevLett.112.232301} {\bibfield  {journal} {\bibinfo  {journal}
  {Phys. Rev. Lett.}\ }\textbf {\bibinfo {volume} {112}},\ \bibinfo {pages}
  {232301} (\bibinfo {year} {2014})},\ \Eprint {http://arxiv.org/abs/1307.4379}
  {arXiv:1307.4379 [nucl-th]} \BibitemShut {NoStop}%
\bibitem [{\citenamefont {Kozlov}\ \emph {et~al.}(2014)\citenamefont {Kozlov},
  \citenamefont {Luzum}, \citenamefont {Denicol}, \citenamefont {Jeon},\ and\
  \citenamefont {Gale}}]{Kozlov:2014fqa}%
  \BibitemOpen
  \bibfield  {author} {\bibinfo {author} {\bibfnamefont {I.}~\bibnamefont
  {Kozlov}}, \bibinfo {author} {\bibfnamefont {M.}~\bibnamefont {Luzum}},
  \bibinfo {author} {\bibfnamefont {G.}~\bibnamefont {Denicol}}, \bibinfo
  {author} {\bibfnamefont {S.}~\bibnamefont {Jeon}}, \ and\ \bibinfo {author}
  {\bibfnamefont {C.}~\bibnamefont {Gale}},\ }\href@noop {} {\  (\bibinfo
  {year} {2014})},\ \Eprint {http://arxiv.org/abs/1405.3976} {arXiv:1405.3976
  [nucl-th]} \BibitemShut {NoStop}%
\bibitem [{\citenamefont {Schenke}\ and\ \citenamefont
  {Venugopalan}(2014)}]{Schenke:2014zha}%
  \BibitemOpen
  \bibfield  {author} {\bibinfo {author} {\bibfnamefont {B.}~\bibnamefont
  {Schenke}}\ and\ \bibinfo {author} {\bibfnamefont {R.}~\bibnamefont
  {Venugopalan}},\ }\href {\doibase 10.1103/PhysRevLett.113.102301} {\bibfield
  {journal} {\bibinfo  {journal} {Phys.Rev.Lett.}\ }\textbf {\bibinfo {volume}
  {113}},\ \bibinfo {pages} {102301} (\bibinfo {year} {2014})},\ \Eprint
  {http://arxiv.org/abs/1405.3605} {arXiv:1405.3605 [nucl-th]} \BibitemShut
  {NoStop}%
\bibitem [{\citenamefont {Romatschke}(2015)}]{Romatschke:2015gxa}%
  \BibitemOpen
  \bibfield  {author} {\bibinfo {author} {\bibfnamefont {P.}~\bibnamefont
  {Romatschke}},\ }\href {\doibase 10.1140/epjc/s10052-015-3509-3} {\bibfield
  {journal} {\bibinfo  {journal} {Eur. Phys. J.}\ }\textbf {\bibinfo {volume}
  {C75}},\ \bibinfo {pages} {305} (\bibinfo {year} {2015})},\ \Eprint
  {http://arxiv.org/abs/1502.04745} {arXiv:1502.04745 [nucl-th]} \BibitemShut
  {NoStop}%
\bibitem [{\citenamefont {Shen}\ \emph {et~al.}(2017)\citenamefont {Shen},
  \citenamefont {Paquet}, \citenamefont {Denicol}, \citenamefont {Jeon},\ and\
  \citenamefont {Gale}}]{Shen:2016zpp}%
  \BibitemOpen
  \bibfield  {author} {\bibinfo {author} {\bibfnamefont {C.}~\bibnamefont
  {Shen}}, \bibinfo {author} {\bibfnamefont {J.-F.}\ \bibnamefont {Paquet}},
  \bibinfo {author} {\bibfnamefont {G.~S.}\ \bibnamefont {Denicol}}, \bibinfo
  {author} {\bibfnamefont {S.}~\bibnamefont {Jeon}}, \ and\ \bibinfo {author}
  {\bibfnamefont {C.}~\bibnamefont {Gale}},\ }\href {\doibase
  10.1103/PhysRevC.95.014906} {\bibfield  {journal} {\bibinfo  {journal} {Phys.
  Rev.}\ }\textbf {\bibinfo {volume} {C95}},\ \bibinfo {pages} {014906}
  (\bibinfo {year} {2017})},\ \Eprint {http://arxiv.org/abs/1609.02590}
  {arXiv:1609.02590 [nucl-th]} \BibitemShut {NoStop}%
\bibitem [{\citenamefont {Weller}\ and\ \citenamefont
  {Romatschke}(2017)}]{Weller:2017tsr}%
  \BibitemOpen
  \bibfield  {author} {\bibinfo {author} {\bibfnamefont {R.~D.}\ \bibnamefont
  {Weller}}\ and\ \bibinfo {author} {\bibfnamefont {P.}~\bibnamefont
  {Romatschke}},\ }\href@noop {} {\  (\bibinfo {year} {2017})},\ \Eprint
  {http://arxiv.org/abs/1701.07145} {arXiv:1701.07145 [nucl-th]} \BibitemShut
  {NoStop}%
\bibitem [{\citenamefont {Mäntysaari}\ \emph {et~al.}(2017)\citenamefont
  {Mäntysaari}, \citenamefont {Schenke}, \citenamefont {Shen},\ and\
  \citenamefont {Tribedy}}]{Mantysaari:2017cni}%
  \BibitemOpen
  \bibfield  {author} {\bibinfo {author} {\bibfnamefont {H.}~\bibnamefont
  {Mäntysaari}}, \bibinfo {author} {\bibfnamefont {B.}~\bibnamefont
  {Schenke}}, \bibinfo {author} {\bibfnamefont {C.}~\bibnamefont {Shen}}, \
  and\ \bibinfo {author} {\bibfnamefont {P.}~\bibnamefont {Tribedy}},\ }\href
  {\doibase 10.1016/j.physletb.2017.07.038} {\bibfield  {journal} {\bibinfo
  {journal} {Phys. Lett.}\ }\textbf {\bibinfo {volume} {B772}},\ \bibinfo
  {pages} {681} (\bibinfo {year} {2017})},\ \Eprint
  {http://arxiv.org/abs/1705.03177} {arXiv:1705.03177 [nucl-th]} \BibitemShut
  {NoStop}%
\bibitem [{\citenamefont {Schenke}\ \emph
  {et~al.}(2012{\natexlab{a}})\citenamefont {Schenke}, \citenamefont
  {Tribedy},\ and\ \citenamefont {Venugopalan}}]{Schenke:2012wb}%
  \BibitemOpen
  \bibfield  {author} {\bibinfo {author} {\bibfnamefont {B.}~\bibnamefont
  {Schenke}}, \bibinfo {author} {\bibfnamefont {P.}~\bibnamefont {Tribedy}}, \
  and\ \bibinfo {author} {\bibfnamefont {R.}~\bibnamefont {Venugopalan}},\
  }\href {\doibase 10.1103/PhysRevLett.108.252301} {\bibfield  {journal}
  {\bibinfo  {journal} {Phys. Rev. Lett.}\ }\textbf {\bibinfo {volume} {108}},\
  \bibinfo {pages} {252301} (\bibinfo {year} {2012}{\natexlab{a}})},\ \Eprint
  {http://arxiv.org/abs/1202.6646} {arXiv:1202.6646 [nucl-th]} \BibitemShut
  {NoStop}%
\bibitem [{\citenamefont {Schenke}\ \emph
  {et~al.}(2012{\natexlab{b}})\citenamefont {Schenke}, \citenamefont
  {Tribedy},\ and\ \citenamefont {Venugopalan}}]{Schenke:2012hg}%
  \BibitemOpen
  \bibfield  {author} {\bibinfo {author} {\bibfnamefont {B.}~\bibnamefont
  {Schenke}}, \bibinfo {author} {\bibfnamefont {P.}~\bibnamefont {Tribedy}}, \
  and\ \bibinfo {author} {\bibfnamefont {R.}~\bibnamefont {Venugopalan}},\
  }\href {\doibase 10.1103/PhysRevC.86.034908} {\bibfield  {journal} {\bibinfo
  {journal} {Phys. Rev.}\ }\textbf {\bibinfo {volume} {C86}},\ \bibinfo {pages}
  {034908} (\bibinfo {year} {2012}{\natexlab{b}})},\ \Eprint
  {http://arxiv.org/abs/1206.6805} {arXiv:1206.6805 [hep-ph]} \BibitemShut
  {NoStop}%
\bibitem [{\citenamefont {Greif}\ \emph {et~al.}(2017)\citenamefont {Greif},
  \citenamefont {Greiner}, \citenamefont {Schenke}, \citenamefont
  {Schlichting},\ and\ \citenamefont {Xu}}]{Greif:2017bnr}%
  \BibitemOpen
  \bibfield  {author} {\bibinfo {author} {\bibfnamefont {M.}~\bibnamefont
  {Greif}}, \bibinfo {author} {\bibfnamefont {C.}~\bibnamefont {Greiner}},
  \bibinfo {author} {\bibfnamefont {B.}~\bibnamefont {Schenke}}, \bibinfo
  {author} {\bibfnamefont {S.}~\bibnamefont {Schlichting}}, \ and\ \bibinfo
  {author} {\bibfnamefont {Z.}~\bibnamefont {Xu}},\ }\href {\doibase
  10.1103/PhysRevD.96.091504} {\bibfield  {journal} {\bibinfo  {journal} {Phys.
  Rev.}\ }\textbf {\bibinfo {volume} {D96}},\ \bibinfo {pages} {091504}
  (\bibinfo {year} {2017})},\ \Eprint {http://arxiv.org/abs/1708.02076}
  {arXiv:1708.02076 [hep-ph]} \BibitemShut {NoStop}%
\bibitem [{\citenamefont {Schenke}\ \emph {et~al.}(2010)\citenamefont
  {Schenke}, \citenamefont {Jeon},\ and\ \citenamefont
  {Gale}}]{Schenke:2010nt}%
  \BibitemOpen
  \bibfield  {author} {\bibinfo {author} {\bibfnamefont {B.}~\bibnamefont
  {Schenke}}, \bibinfo {author} {\bibfnamefont {S.}~\bibnamefont {Jeon}}, \
  and\ \bibinfo {author} {\bibfnamefont {C.}~\bibnamefont {Gale}},\ }\href
  {\doibase 10.1103/PhysRevC.82.014903} {\bibfield  {journal} {\bibinfo
  {journal} {Phys. Rev.}\ }\textbf {\bibinfo {volume} {C82}},\ \bibinfo {pages}
  {014903} (\bibinfo {year} {2010})},\ \Eprint {http://arxiv.org/abs/1004.1408}
  {arXiv:1004.1408 [hep-ph]} \BibitemShut {NoStop}%
\bibitem [{\citenamefont {Schenke}\ \emph {et~al.}(2011)\citenamefont
  {Schenke}, \citenamefont {Jeon},\ and\ \citenamefont
  {Gale}}]{Schenke:2010rr}%
  \BibitemOpen
  \bibfield  {author} {\bibinfo {author} {\bibfnamefont {B.}~\bibnamefont
  {Schenke}}, \bibinfo {author} {\bibfnamefont {S.}~\bibnamefont {Jeon}}, \
  and\ \bibinfo {author} {\bibfnamefont {C.}~\bibnamefont {Gale}},\ }\href
  {\doibase 10.1103/PhysRevLett.106.042301} {\bibfield  {journal} {\bibinfo
  {journal} {Phys. Rev. Lett.}\ }\textbf {\bibinfo {volume} {106}},\ \bibinfo
  {pages} {042301} (\bibinfo {year} {2011})},\ \Eprint
  {http://arxiv.org/abs/1009.3244} {arXiv:1009.3244 [hep-ph]} \BibitemShut
  {NoStop}%
\bibitem [{\citenamefont {Schenke}\ \emph
  {et~al.}(2012{\natexlab{c}})\citenamefont {Schenke}, \citenamefont {Jeon},\
  and\ \citenamefont {Gale}}]{Schenke:2011bn}%
  \BibitemOpen
  \bibfield  {author} {\bibinfo {author} {\bibfnamefont {B.}~\bibnamefont
  {Schenke}}, \bibinfo {author} {\bibfnamefont {S.}~\bibnamefont {Jeon}}, \
  and\ \bibinfo {author} {\bibfnamefont {C.}~\bibnamefont {Gale}},\ }\href
  {\doibase 10.1103/PhysRevC.85.024901} {\bibfield  {journal} {\bibinfo
  {journal} {Phys. Rev.}\ }\textbf {\bibinfo {volume} {C85}},\ \bibinfo {pages}
  {024901} (\bibinfo {year} {2012}{\natexlab{c}})},\ \Eprint
  {http://arxiv.org/abs/1109.6289} {arXiv:1109.6289 [hep-ph]} \BibitemShut
  {NoStop}%
\bibitem [{\citenamefont {Aidala}\ \emph {et~al.}(2019)\citenamefont {Aidala}
  \emph {et~al.}}]{PHENIX:2018lia}%
  \BibitemOpen
  \bibfield  {author} {\bibinfo {author} {\bibfnamefont {C.}~\bibnamefont
  {Aidala}} \emph {et~al.} (\bibinfo {collaboration} {PHENIX}),\ }\href
  {\doibase 10.1038/s41567-018-0360-0} {\bibfield  {journal} {\bibinfo
  {journal} {Nature Phys.}\ }\textbf {\bibinfo {volume} {15}},\ \bibinfo
  {pages} {214} (\bibinfo {year} {2019})},\ \Eprint
  {http://arxiv.org/abs/1805.02973} {arXiv:1805.02973 [nucl-ex]} \BibitemShut
  {NoStop}%
\bibitem [{\citenamefont {Bass}\ \emph {et~al.}(1998)\citenamefont {Bass} \emph
  {et~al.}}]{Bass:1998ca}%
  \BibitemOpen
  \bibfield  {author} {\bibinfo {author} {\bibfnamefont {S.~A.}\ \bibnamefont
  {Bass}} \emph {et~al.},\ }\href {\doibase 10.1016/S0146-6410(98)00058-1}
  {\bibfield  {journal} {\bibinfo  {journal} {Prog. Part. Nucl. Phys.}\
  }\textbf {\bibinfo {volume} {41}},\ \bibinfo {pages} {255} (\bibinfo {year}
  {1998})},\ \bibinfo {note} {[Prog. Part. Nucl. Phys.41,225(1998)]},\ \Eprint
  {http://arxiv.org/abs/nucl-th/9803035} {arXiv:nucl-th/9803035 [nucl-th]}
  \BibitemShut {NoStop}%
\bibitem [{\citenamefont {Bleicher}\ \emph {et~al.}(1999)\citenamefont
  {Bleicher} \emph {et~al.}}]{Bleicher:1999xi}%
  \BibitemOpen
  \bibfield  {author} {\bibinfo {author} {\bibfnamefont {M.}~\bibnamefont
  {Bleicher}} \emph {et~al.},\ }\href {\doibase 10.1088/0954-3899/25/9/308}
  {\bibfield  {journal} {\bibinfo  {journal} {J. Phys.}\ }\textbf {\bibinfo
  {volume} {G25}},\ \bibinfo {pages} {1859} (\bibinfo {year} {1999})},\ \Eprint
  {http://arxiv.org/abs/hep-ph/9909407} {arXiv:hep-ph/9909407 [hep-ph]}
  \BibitemShut {NoStop}%
\bibitem [{\citenamefont {Schenke}\ \emph {et~al.}(2019)\citenamefont
  {Schenke}, \citenamefont {Shen},\ and\ \citenamefont
  {Tribedy}}]{Schenke:2019ruo}%
  \BibitemOpen
  \bibfield  {author} {\bibinfo {author} {\bibfnamefont {B.}~\bibnamefont
  {Schenke}}, \bibinfo {author} {\bibfnamefont {C.}~\bibnamefont {Shen}}, \
  and\ \bibinfo {author} {\bibfnamefont {P.}~\bibnamefont {Tribedy}},\ }\href
  {\doibase 10.1103/PhysRevC.99.044908} {\bibfield  {journal} {\bibinfo
  {journal} {Phys. Rev.}\ }\textbf {\bibinfo {volume} {C99}},\ \bibinfo {pages}
  {044908} (\bibinfo {year} {2019})},\ \Eprint
  {http://arxiv.org/abs/1901.04378} {arXiv:1901.04378 [nucl-th]} \BibitemShut
  {NoStop}%
\bibitem [{\citenamefont {Nagle}\ and\ \citenamefont
  {Zajc}(2018)}]{Nagle:2018ybc}%
  \BibitemOpen
  \bibfield  {author} {\bibinfo {author} {\bibfnamefont {J.~L.}\ \bibnamefont
  {Nagle}}\ and\ \bibinfo {author} {\bibfnamefont {W.~A.}\ \bibnamefont
  {Zajc}},\ }\href@noop {} {\  (\bibinfo {year} {2018})},\ \Eprint
  {http://arxiv.org/abs/1808.01276} {arXiv:1808.01276 [nucl-th]} \BibitemShut
  {NoStop}%
\bibitem [{\citenamefont {Lim}\ \emph {et~al.}(2018)\citenamefont {Lim},
  \citenamefont {Carlson}, \citenamefont {Loizides}, \citenamefont {Lonardoni},
  \citenamefont {Lynn}, \citenamefont {Nagle}, \citenamefont {Orjuela~Koop},\
  and\ \citenamefont {Ouellette}}]{Lim:2018huo}%
  \BibitemOpen
  \bibfield  {author} {\bibinfo {author} {\bibfnamefont {S.~H.}\ \bibnamefont
  {Lim}}, \bibinfo {author} {\bibfnamefont {J.}~\bibnamefont {Carlson}},
  \bibinfo {author} {\bibfnamefont {C.}~\bibnamefont {Loizides}}, \bibinfo
  {author} {\bibfnamefont {D.}~\bibnamefont {Lonardoni}}, \bibinfo {author}
  {\bibfnamefont {J.~E.}\ \bibnamefont {Lynn}}, \bibinfo {author}
  {\bibfnamefont {J.~L.}\ \bibnamefont {Nagle}}, \bibinfo {author}
  {\bibfnamefont {J.~D.}\ \bibnamefont {Orjuela~Koop}}, \ and\ \bibinfo
  {author} {\bibfnamefont {J.}~\bibnamefont {Ouellette}},\ }\href@noop {} {\
  (\bibinfo {year} {2018})},\ \Eprint {http://arxiv.org/abs/1812.08096}
  {arXiv:1812.08096 [nucl-th]} \BibitemShut {NoStop}%
\bibitem [{\citenamefont {Sievert}\ and\ \citenamefont
  {Noronha-Hostler}(2019)}]{Sievert:2019zjr}%
  \BibitemOpen
  \bibfield  {author} {\bibinfo {author} {\bibfnamefont {M.~D.}\ \bibnamefont
  {Sievert}}\ and\ \bibinfo {author} {\bibfnamefont {J.}~\bibnamefont
  {Noronha-Hostler}},\ }\href@noop {} {\  (\bibinfo {year} {2019})},\ \Eprint
  {http://arxiv.org/abs/1901.01319} {arXiv:1901.01319 [nucl-th]} \BibitemShut
  {NoStop}%
\bibitem [{\citenamefont {Moreland}\ \emph {et~al.}(2015)\citenamefont
  {Moreland}, \citenamefont {Bernhard},\ and\ \citenamefont
  {Bass}}]{Moreland:2014oya}%
  \BibitemOpen
  \bibfield  {author} {\bibinfo {author} {\bibfnamefont {J.~S.}\ \bibnamefont
  {Moreland}}, \bibinfo {author} {\bibfnamefont {J.~E.}\ \bibnamefont
  {Bernhard}}, \ and\ \bibinfo {author} {\bibfnamefont {S.~A.}\ \bibnamefont
  {Bass}},\ }\href {\doibase 10.1103/PhysRevC.92.011901} {\bibfield  {journal}
  {\bibinfo  {journal} {Phys. Rev.}\ }\textbf {\bibinfo {volume} {C92}},\
  \bibinfo {pages} {011901} (\bibinfo {year} {2015})},\ \Eprint
  {http://arxiv.org/abs/1412.4708} {arXiv:1412.4708 [nucl-th]} \BibitemShut
  {NoStop}%
\bibitem [{\citenamefont {Hulthen}\ and\ \citenamefont
  {Sagawara}(1957)}]{Hulthen:1957}%
  \BibitemOpen
  \bibfield  {author} {\bibinfo {author} {\bibfnamefont {L.}~\bibnamefont
  {Hulthen}}\ and\ \bibinfo {author} {\bibfnamefont {M.}~\bibnamefont
  {Sagawara}},\ }\href@noop {} {\emph {\bibinfo {title} {Handbuch der Physik
  39, 14}}}\ (\bibinfo {year} {1957})\BibitemShut {NoStop}%
\bibitem [{\citenamefont {Miller}\ \emph {et~al.}(2007)\citenamefont {Miller},
  \citenamefont {Reygers}, \citenamefont {Sanders},\ and\ \citenamefont
  {Steinberg}}]{Miller:2007ri}%
  \BibitemOpen
  \bibfield  {author} {\bibinfo {author} {\bibfnamefont {M.~L.}\ \bibnamefont
  {Miller}}, \bibinfo {author} {\bibfnamefont {K.}~\bibnamefont {Reygers}},
  \bibinfo {author} {\bibfnamefont {S.~J.}\ \bibnamefont {Sanders}}, \ and\
  \bibinfo {author} {\bibfnamefont {P.}~\bibnamefont {Steinberg}},\ }\href
  {\doibase 10.1146/annurev.nucl.57.090506.123020} {\bibfield  {journal}
  {\bibinfo  {journal} {Ann. Rev. Nucl. Part. Sci.}\ }\textbf {\bibinfo
  {volume} {57}},\ \bibinfo {pages} {205} (\bibinfo {year} {2007})}\BibitemShut
  {NoStop}%
\bibitem [{\citenamefont {Nagle}\ \emph {et~al.}(2014)\citenamefont {Nagle},
  \citenamefont {Adare}, \citenamefont {Beckman}, \citenamefont {Koblesky},
  \citenamefont {Orjuela~Koop}, \citenamefont {McGlinchey}, \citenamefont
  {Romatschke}, \citenamefont {Carlson}, \citenamefont {Lynn},\ and\
  \citenamefont {McCumber}}]{Nagle:2013lja}%
  \BibitemOpen
  \bibfield  {author} {\bibinfo {author} {\bibfnamefont {J.~L.}\ \bibnamefont
  {Nagle}}, \bibinfo {author} {\bibfnamefont {A.}~\bibnamefont {Adare}},
  \bibinfo {author} {\bibfnamefont {S.}~\bibnamefont {Beckman}}, \bibinfo
  {author} {\bibfnamefont {T.}~\bibnamefont {Koblesky}}, \bibinfo {author}
  {\bibfnamefont {J.}~\bibnamefont {Orjuela~Koop}}, \bibinfo {author}
  {\bibfnamefont {D.}~\bibnamefont {McGlinchey}}, \bibinfo {author}
  {\bibfnamefont {P.}~\bibnamefont {Romatschke}}, \bibinfo {author}
  {\bibfnamefont {J.}~\bibnamefont {Carlson}}, \bibinfo {author} {\bibfnamefont
  {J.~E.}\ \bibnamefont {Lynn}}, \ and\ \bibinfo {author} {\bibfnamefont
  {M.}~\bibnamefont {McCumber}},\ }\href {\doibase
  10.1103/PhysRevLett.113.112301} {\bibfield  {journal} {\bibinfo  {journal}
  {Phys. Rev. Lett.}\ }\textbf {\bibinfo {volume} {113}},\ \bibinfo {pages}
  {112301} (\bibinfo {year} {2014})},\ \Eprint {http://arxiv.org/abs/1312.4565}
  {arXiv:1312.4565 [nucl-th]} \BibitemShut {NoStop}%
\bibitem [{\citenamefont {Carlson}\ and\ \citenamefont
  {Schiavilla}(1998)}]{Carlson:1997qn}%
  \BibitemOpen
  \bibfield  {author} {\bibinfo {author} {\bibfnamefont {J.}~\bibnamefont
  {Carlson}}\ and\ \bibinfo {author} {\bibfnamefont {R.}~\bibnamefont
  {Schiavilla}},\ }\href {\doibase 10.1103/RevModPhys.70.743} {\bibfield
  {journal} {\bibinfo  {journal} {Rev. Mod. Phys.}\ }\textbf {\bibinfo {volume}
  {70}},\ \bibinfo {pages} {743} (\bibinfo {year} {1998})}\BibitemShut
  {NoStop}%
\bibitem [{\citenamefont {Kowalski}\ and\ \citenamefont
  {Teaney}(2003)}]{Kowalski:2003hm}%
  \BibitemOpen
  \bibfield  {author} {\bibinfo {author} {\bibfnamefont {H.}~\bibnamefont
  {Kowalski}}\ and\ \bibinfo {author} {\bibfnamefont {D.}~\bibnamefont
  {Teaney}},\ }\href {\doibase 10.1103/PhysRevD.68.114005} {\bibfield
  {journal} {\bibinfo  {journal} {Phys. Rev.}\ }\textbf {\bibinfo {volume}
  {D68}},\ \bibinfo {pages} {114005} (\bibinfo {year} {2003})}\BibitemShut
  {NoStop}%
\bibitem [{\citenamefont {Rezaeian}\ \emph {et~al.}(2013)\citenamefont
  {Rezaeian}, \citenamefont {Siddikov}, \citenamefont {Van~de Klundert},\ and\
  \citenamefont {Venugopalan}}]{Rezaeian:2012ji}%
  \BibitemOpen
  \bibfield  {author} {\bibinfo {author} {\bibfnamefont {A.~H.}\ \bibnamefont
  {Rezaeian}}, \bibinfo {author} {\bibfnamefont {M.}~\bibnamefont {Siddikov}},
  \bibinfo {author} {\bibfnamefont {M.}~\bibnamefont {Van~de Klundert}}, \ and\
  \bibinfo {author} {\bibfnamefont {R.}~\bibnamefont {Venugopalan}},\
  }\href@noop {} {\bibfield  {journal} {\bibinfo  {journal} {Phys.Rev.}\
  }\textbf {\bibinfo {volume} {D87}},\ \bibinfo {pages} {034002} (\bibinfo
  {year} {2013})},\ \Eprint {http://arxiv.org/abs/1212.2974} {arXiv:1212.2974
  [hep-ph]} \BibitemShut {NoStop}%
\bibitem [{\citenamefont {Krasnitz}\ and\ \citenamefont
  {Venugopalan}(1999)}]{Krasnitz:1998ns}%
  \BibitemOpen
  \bibfield  {author} {\bibinfo {author} {\bibfnamefont {A.}~\bibnamefont
  {Krasnitz}}\ and\ \bibinfo {author} {\bibfnamefont {R.}~\bibnamefont
  {Venugopalan}},\ }\href@noop {} {\bibfield  {journal} {\bibinfo  {journal}
  {Nucl. Phys.}\ }\textbf {\bibinfo {volume} {B557}},\ \bibinfo {pages} {237}
  (\bibinfo {year} {1999})}\BibitemShut {NoStop}%
\bibitem [{\citenamefont {Bazavov}\ \emph {et~al.}(2014)\citenamefont {Bazavov}
  \emph {et~al.}}]{Bazavov:2014pvz}%
  \BibitemOpen
  \bibfield  {author} {\bibinfo {author} {\bibfnamefont {A.}~\bibnamefont
  {Bazavov}} \emph {et~al.} (\bibinfo {collaboration} {HotQCD}),\ }\href
  {\doibase 10.1103/PhysRevD.90.094503} {\bibfield  {journal} {\bibinfo
  {journal} {Phys. Rev.}\ }\textbf {\bibinfo {volume} {D90}},\ \bibinfo {pages}
  {094503} (\bibinfo {year} {2014})},\ \Eprint {http://arxiv.org/abs/1407.6387}
  {arXiv:1407.6387 [hep-lat]} \BibitemShut {NoStop}%
\bibitem [{\citenamefont {Moreland}\ and\ \citenamefont
  {Soltz}(2016)}]{Moreland:2015dvc}%
  \BibitemOpen
  \bibfield  {author} {\bibinfo {author} {\bibfnamefont {J.~S.}\ \bibnamefont
  {Moreland}}\ and\ \bibinfo {author} {\bibfnamefont {R.~A.}\ \bibnamefont
  {Soltz}},\ }\href {\doibase 10.1103/PhysRevC.93.044913} {\bibfield  {journal}
  {\bibinfo  {journal} {Phys. Rev.}\ }\textbf {\bibinfo {volume} {C93}},\
  \bibinfo {pages} {044913} (\bibinfo {year} {2016})},\ \Eprint
  {http://arxiv.org/abs/1512.02189} {arXiv:1512.02189 [nucl-th]} \BibitemShut
  {NoStop}%
\bibitem [{\citenamefont {Denicol}\ \emph {et~al.}(2012)\citenamefont
  {Denicol}, \citenamefont {Niemi}, \citenamefont {Molnar},\ and\ \citenamefont
  {Rischke}}]{Denicol:2012cn}%
  \BibitemOpen
  \bibfield  {author} {\bibinfo {author} {\bibfnamefont {G.~S.}\ \bibnamefont
  {Denicol}}, \bibinfo {author} {\bibfnamefont {H.}~\bibnamefont {Niemi}},
  \bibinfo {author} {\bibfnamefont {E.}~\bibnamefont {Molnar}}, \ and\ \bibinfo
  {author} {\bibfnamefont {D.~H.}\ \bibnamefont {Rischke}},\ }\href {\doibase
  10.1103/PhysRevD.85.114047, 10.1103/PhysRevD.91.039902} {\bibfield  {journal}
  {\bibinfo  {journal} {Phys. Rev.}\ }\textbf {\bibinfo {volume} {D85}},\
  \bibinfo {pages} {114047} (\bibinfo {year} {2012})},\ \bibinfo {note}
  {[Erratum: Phys. Rev.D91,no.3,039902(2015)]},\ \Eprint
  {http://arxiv.org/abs/1202.4551} {arXiv:1202.4551 [nucl-th]} \BibitemShut
  {NoStop}%
\bibitem [{\citenamefont {Denicol}\ \emph {et~al.}(2014)\citenamefont
  {Denicol}, \citenamefont {Jeon},\ and\ \citenamefont
  {Gale}}]{Denicol:2014vaa}%
  \BibitemOpen
  \bibfield  {author} {\bibinfo {author} {\bibfnamefont {G.~S.}\ \bibnamefont
  {Denicol}}, \bibinfo {author} {\bibfnamefont {S.}~\bibnamefont {Jeon}}, \
  and\ \bibinfo {author} {\bibfnamefont {C.}~\bibnamefont {Gale}},\ }\href
  {\doibase 10.1103/PhysRevC.90.024912} {\bibfield  {journal} {\bibinfo
  {journal} {Phys. Rev.}\ }\textbf {\bibinfo {volume} {C90}},\ \bibinfo {pages}
  {024912} (\bibinfo {year} {2014})},\ \Eprint {http://arxiv.org/abs/1403.0962}
  {arXiv:1403.0962 [nucl-th]} \BibitemShut {NoStop}%
\bibitem [{\citenamefont {Cooper}\ and\ \citenamefont
  {Frye}(1974)}]{Cooper:1974mv}%
  \BibitemOpen
  \bibfield  {author} {\bibinfo {author} {\bibfnamefont {F.}~\bibnamefont
  {Cooper}}\ and\ \bibinfo {author} {\bibfnamefont {G.}~\bibnamefont {Frye}},\
  }\href {\doibase 10.1103/PhysRevD.10.186} {\bibfield  {journal} {\bibinfo
  {journal} {Phys. Rev.}\ }\textbf {\bibinfo {volume} {D10}},\ \bibinfo {pages}
  {186} (\bibinfo {year} {1974})}\BibitemShut {NoStop}%
\bibitem [{\citenamefont {Dusling}\ \emph {et~al.}(2010)\citenamefont
  {Dusling}, \citenamefont {Moore},\ and\ \citenamefont
  {Teaney}}]{Dusling:2009df}%
  \BibitemOpen
  \bibfield  {author} {\bibinfo {author} {\bibfnamefont {K.}~\bibnamefont
  {Dusling}}, \bibinfo {author} {\bibfnamefont {G.~D.}\ \bibnamefont {Moore}},
  \ and\ \bibinfo {author} {\bibfnamefont {D.}~\bibnamefont {Teaney}},\ }\href
  {\doibase 10.1103/PhysRevC.81.034907} {\bibfield  {journal} {\bibinfo
  {journal} {Phys. Rev.}\ }\textbf {\bibinfo {volume} {C81}},\ \bibinfo {pages}
  {034907} (\bibinfo {year} {2010})}\BibitemShut {NoStop}%
\bibitem [{\citenamefont {Bozek}(2010)}]{Bozek:2009dw}%
  \BibitemOpen
  \bibfield  {author} {\bibinfo {author} {\bibfnamefont {P.}~\bibnamefont
  {Bozek}},\ }\href {\doibase 10.1103/PhysRevC.81.034909} {\bibfield  {journal}
  {\bibinfo  {journal} {Phys. Rev.}\ }\textbf {\bibinfo {volume} {C81}},\
  \bibinfo {pages} {034909} (\bibinfo {year} {2010})},\ \Eprint
  {http://arxiv.org/abs/0911.2397} {arXiv:0911.2397 [nucl-th]} \BibitemShut
  {NoStop}%
\bibitem [{\citenamefont {Paquet}\ \emph {et~al.}(2016)\citenamefont {Paquet},
  \citenamefont {Shen}, \citenamefont {Denicol}, \citenamefont {Luzum},
  \citenamefont {Schenke}, \citenamefont {Jeon},\ and\ \citenamefont
  {Gale}}]{Paquet:2015lta}%
  \BibitemOpen
  \bibfield  {author} {\bibinfo {author} {\bibfnamefont {J.-F.}\ \bibnamefont
  {Paquet}}, \bibinfo {author} {\bibfnamefont {C.}~\bibnamefont {Shen}},
  \bibinfo {author} {\bibfnamefont {G.~S.}\ \bibnamefont {Denicol}}, \bibinfo
  {author} {\bibfnamefont {M.}~\bibnamefont {Luzum}}, \bibinfo {author}
  {\bibfnamefont {B.}~\bibnamefont {Schenke}}, \bibinfo {author} {\bibfnamefont
  {S.}~\bibnamefont {Jeon}}, \ and\ \bibinfo {author} {\bibfnamefont
  {C.}~\bibnamefont {Gale}},\ }\href {\doibase 10.1103/PhysRevC.93.044906}
  {\bibfield  {journal} {\bibinfo  {journal} {Phys. Rev.}\ }\textbf {\bibinfo
  {volume} {C93}},\ \bibinfo {pages} {044906} (\bibinfo {year} {2016})},\
  \Eprint {http://arxiv.org/abs/1509.06738} {arXiv:1509.06738 [hep-ph]}
  \BibitemShut {NoStop}%
\bibitem [{\citenamefont {Abelev}\ \emph {et~al.}(2009)\citenamefont {Abelev}
  \emph {et~al.}}]{Abelev:2008ab}%
  \BibitemOpen
  \bibfield  {author} {\bibinfo {author} {\bibfnamefont {B.~I.}\ \bibnamefont
  {Abelev}} \emph {et~al.} (\bibinfo {collaboration} {STAR}),\ }\href {\doibase
  10.1103/PhysRevC.79.034909} {\bibfield  {journal} {\bibinfo  {journal} {Phys.
  Rev.}\ }\textbf {\bibinfo {volume} {C79}},\ \bibinfo {pages} {034909}
  (\bibinfo {year} {2009})},\ \Eprint {http://arxiv.org/abs/0808.2041}
  {arXiv:0808.2041 [nucl-ex]} \BibitemShut {NoStop}%
\bibitem [{\citenamefont {Ansorge}\ \emph {et~al.}(1989)\citenamefont {Ansorge}
  \emph {et~al.}}]{Ansorge1989}%
  \BibitemOpen
  \bibfield  {author} {\bibinfo {author} {\bibfnamefont {R.}~\bibnamefont
  {Ansorge}} \emph {et~al.} (\bibinfo {collaboration} {UA5 Collaboration}),\
  }\href {\doibase 10.1007/BF01506531} {\bibfield  {journal} {\bibinfo
  {journal} {Zeitschrift f{\"u}r Physik C Particles and Fields}\ }\textbf
  {\bibinfo {volume} {43}},\ \bibinfo {pages} {357} (\bibinfo {year}
  {1989})}\BibitemShut {NoStop}%
\bibitem [{\citenamefont {Aidala}\ \emph {et~al.}(2018)\citenamefont {Aidala}
  \emph {et~al.}}]{Aidala:2017ajz}%
  \BibitemOpen
  \bibfield  {author} {\bibinfo {author} {\bibfnamefont {C.}~\bibnamefont
  {Aidala}} \emph {et~al.} (\bibinfo {collaboration} {PHENIX}),\ }\href
  {\doibase 10.1103/PhysRevLett.120.062302} {\bibfield  {journal} {\bibinfo
  {journal} {Phys. Rev. Lett.}\ }\textbf {\bibinfo {volume} {120}},\ \bibinfo
  {pages} {062302} (\bibinfo {year} {2018})},\ \Eprint
  {http://arxiv.org/abs/1707.06108} {arXiv:1707.06108 [nucl-ex]} \BibitemShut
  {NoStop}%
\bibitem [{\citenamefont {Luzum}\ and\ \citenamefont
  {Ollitrault}(2013)}]{Luzum:2012da}%
  \BibitemOpen
  \bibfield  {author} {\bibinfo {author} {\bibfnamefont {M.}~\bibnamefont
  {Luzum}}\ and\ \bibinfo {author} {\bibfnamefont {J.-Y.}\ \bibnamefont
  {Ollitrault}},\ }\href {\doibase 10.1103/PhysRevC.87.044907} {\bibfield
  {journal} {\bibinfo  {journal} {Phys. Rev.}\ }\textbf {\bibinfo {volume}
  {C87}},\ \bibinfo {pages} {044907} (\bibinfo {year} {2013})},\ \Eprint
  {http://arxiv.org/abs/1209.2323} {arXiv:1209.2323 [nucl-ex]} \BibitemShut
  {NoStop}%
\bibitem [{\citenamefont {Mäntysaari}\ and\ \citenamefont
  {Schenke}(2016{\natexlab{a}})}]{Mantysaari:2016ykx}%
  \BibitemOpen
  \bibfield  {author} {\bibinfo {author} {\bibfnamefont {H.}~\bibnamefont
  {Mäntysaari}}\ and\ \bibinfo {author} {\bibfnamefont {B.}~\bibnamefont
  {Schenke}},\ }\href {\doibase 10.1103/PhysRevLett.117.052301} {\bibfield
  {journal} {\bibinfo  {journal} {Phys. Rev. Lett.}\ }\textbf {\bibinfo
  {volume} {117}},\ \bibinfo {pages} {052301} (\bibinfo {year}
  {2016}{\natexlab{a}})},\ \Eprint {http://arxiv.org/abs/1603.04349}
  {arXiv:1603.04349 [hep-ph]} \BibitemShut {NoStop}%
\bibitem [{\citenamefont {Mäntysaari}\ and\ \citenamefont
  {Schenke}(2016{\natexlab{b}})}]{Mantysaari:2016jaz}%
  \BibitemOpen
  \bibfield  {author} {\bibinfo {author} {\bibfnamefont {H.}~\bibnamefont
  {Mäntysaari}}\ and\ \bibinfo {author} {\bibfnamefont {B.}~\bibnamefont
  {Schenke}},\ }\href {\doibase 10.1103/PhysRevD.94.034042} {\bibfield
  {journal} {\bibinfo  {journal} {Phys. Rev.}\ }\textbf {\bibinfo {volume}
  {D94}},\ \bibinfo {pages} {034042} (\bibinfo {year} {2016}{\natexlab{b}})},\
  \Eprint {http://arxiv.org/abs/1607.01711} {arXiv:1607.01711 [hep-ph]}
  \BibitemShut {NoStop}%
\bibitem [{\citenamefont {Aidala}\ \emph {et~al.}(2017)\citenamefont {Aidala}
  \emph {et~al.}}]{Aidala:2017pup}%
  \BibitemOpen
  \bibfield  {author} {\bibinfo {author} {\bibfnamefont {C.}~\bibnamefont
  {Aidala}} \emph {et~al.} (\bibinfo {collaboration} {PHENIX}),\ }\href
  {\doibase 10.1103/PhysRevC.96.064905} {\bibfield  {journal} {\bibinfo
  {journal} {Phys. Rev.}\ }\textbf {\bibinfo {volume} {C96}},\ \bibinfo {pages}
  {064905} (\bibinfo {year} {2017})},\ \Eprint
  {http://arxiv.org/abs/1708.06983} {arXiv:1708.06983 [nucl-ex]} \BibitemShut
  {NoStop}%
\bibitem [{\citenamefont {Krasnitz}\ \emph {et~al.}(2003)\citenamefont
  {Krasnitz}, \citenamefont {Nara},\ and\ \citenamefont
  {Venugopalan}}]{Krasnitz:2002ng}%
  \BibitemOpen
  \bibfield  {author} {\bibinfo {author} {\bibfnamefont {A.}~\bibnamefont
  {Krasnitz}}, \bibinfo {author} {\bibfnamefont {Y.}~\bibnamefont {Nara}}, \
  and\ \bibinfo {author} {\bibfnamefont {R.}~\bibnamefont {Venugopalan}},\
  }\href@noop {} {\bibfield  {journal} {\bibinfo  {journal} {Phys. Lett.}\
  }\textbf {\bibinfo {volume} {B554}},\ \bibinfo {pages} {21} (\bibinfo {year}
  {2003})}\BibitemShut {NoStop}%
\bibitem [{\citenamefont {Adam}\ \emph
  {et~al.}(2019{\natexlab{a}})\citenamefont {Adam} \emph
  {et~al.}}]{Adam:2019woz}%
  \BibitemOpen
  \bibfield  {author} {\bibinfo {author} {\bibfnamefont {J.}~\bibnamefont
  {Adam}} \emph {et~al.} (\bibinfo {collaboration} {STAR}),\ }\href {\doibase
  10.1103/PhysRevLett.122.172301} {\bibfield  {journal} {\bibinfo  {journal}
  {Phys. Rev. Lett.}\ }\textbf {\bibinfo {volume} {122}},\ \bibinfo {pages}
  {172301} (\bibinfo {year} {2019}{\natexlab{a}})},\ \Eprint
  {http://arxiv.org/abs/1901.08155} {arXiv:1901.08155 [nucl-ex]} \BibitemShut
  {NoStop}%
\bibitem [{\citenamefont {Adam}\ \emph
  {et~al.}(2019{\natexlab{b}})\citenamefont {Adam} \emph
  {et~al.}}]{STAR:2019xzd}%
  \BibitemOpen
  \bibfield  {author} {\bibinfo {author} {\bibfnamefont {J.}~\bibnamefont
  {Adam}} \emph {et~al.} (\bibinfo {collaboration} {STAR}),\ }\href@noop {} {\
  (\bibinfo {year} {2019}{\natexlab{b}})},\ \Eprint
  {http://arxiv.org/abs/1906.03373} {arXiv:1906.03373 [nucl-ex]} \BibitemShut
  {NoStop}%
\bibitem [{\citenamefont {Adams}\ \emph {et~al.}(2005)\citenamefont {Adams}
  \emph {et~al.}}]{Adams:2004bi}%
  \BibitemOpen
  \bibfield  {author} {\bibinfo {author} {\bibfnamefont {J.}~\bibnamefont
  {Adams}} \emph {et~al.} (\bibinfo {collaboration} {STAR}),\ }\href {\doibase
  10.1103/PhysRevC.72.014904} {\bibfield  {journal} {\bibinfo  {journal} {Phys.
  Rev.}\ }\textbf {\bibinfo {volume} {C72}},\ \bibinfo {pages} {014904}
  (\bibinfo {year} {2005})}\BibitemShut {NoStop}%
\bibitem [{\citenamefont {Dumitru}\ and\ \citenamefont
  {Skokov}(2015)}]{Dumitru:2014vka}%
  \BibitemOpen
  \bibfield  {author} {\bibinfo {author} {\bibfnamefont {A.}~\bibnamefont
  {Dumitru}}\ and\ \bibinfo {author} {\bibfnamefont {V.}~\bibnamefont
  {Skokov}},\ }\href {\doibase 10.1103/PhysRevD.91.074006} {\bibfield
  {journal} {\bibinfo  {journal} {Phys. Rev.}\ }\textbf {\bibinfo {volume}
  {D91}},\ \bibinfo {pages} {074006} (\bibinfo {year} {2015})},\ \Eprint
  {http://arxiv.org/abs/1411.6630} {arXiv:1411.6630 [hep-ph]} \BibitemShut
  {NoStop}%
\bibitem [{\citenamefont {Skokov}(2015)}]{Skokov:2014tka}%
  \BibitemOpen
  \bibfield  {author} {\bibinfo {author} {\bibfnamefont {V.}~\bibnamefont
  {Skokov}},\ }\href {\doibase 10.1103/PhysRevD.91.054014} {\bibfield
  {journal} {\bibinfo  {journal} {Phys. Rev.}\ }\textbf {\bibinfo {volume}
  {D91}},\ \bibinfo {pages} {054014} (\bibinfo {year} {2015})},\ \Eprint
  {http://arxiv.org/abs/1412.5191} {arXiv:1412.5191 [hep-ph]} \BibitemShut
  {NoStop}%
\bibitem [{\citenamefont {Lappi}\ \emph {et~al.}(2016)\citenamefont {Lappi},
  \citenamefont {Schenke}, \citenamefont {Schlichting},\ and\ \citenamefont
  {Venugopalan}}]{Lappi:2015vta}%
  \BibitemOpen
  \bibfield  {author} {\bibinfo {author} {\bibfnamefont {T.}~\bibnamefont
  {Lappi}}, \bibinfo {author} {\bibfnamefont {B.}~\bibnamefont {Schenke}},
  \bibinfo {author} {\bibfnamefont {S.}~\bibnamefont {Schlichting}}, \ and\
  \bibinfo {author} {\bibfnamefont {R.}~\bibnamefont {Venugopalan}},\ }\href
  {\doibase 10.1007/JHEP01(2016)061} {\bibfield  {journal} {\bibinfo  {journal}
  {JHEP}\ }\textbf {\bibinfo {volume} {01}},\ \bibinfo {pages} {061} (\bibinfo
  {year} {2016})},\ \Eprint {http://arxiv.org/abs/1509.03499} {arXiv:1509.03499
  [hep-ph]} \BibitemShut {NoStop}%
\bibitem [{\citenamefont {Chatrchyan}\ \emph {et~al.}(2013)\citenamefont
  {Chatrchyan} \emph {et~al.}}]{Chatrchyan:2013nka}%
  \BibitemOpen
  \bibfield  {author} {\bibinfo {author} {\bibfnamefont {S.}~\bibnamefont
  {Chatrchyan}} \emph {et~al.} (\bibinfo {collaboration} {CMS Collaboration}),\
  }\href {\doibase 10.1016/j.physletb.2013.06.028} {\bibfield  {journal}
  {\bibinfo  {journal} {Phys.Lett.}\ }\textbf {\bibinfo {volume} {B724}},\
  \bibinfo {pages} {213} (\bibinfo {year} {2013})},\ \Eprint
  {http://arxiv.org/abs/1305.0609} {arXiv:1305.0609 [nucl-ex]} \BibitemShut
  {NoStop}%
\bibitem [{\citenamefont {Abelev}\ \emph {et~al.}(2014)\citenamefont {Abelev}
  \emph {et~al.}}]{Abelev:2014mda}%
  \BibitemOpen
  \bibfield  {author} {\bibinfo {author} {\bibfnamefont {B.~B.}\ \bibnamefont
  {Abelev}} \emph {et~al.} (\bibinfo {collaboration} {ALICE}),\ }\href
  {\doibase 10.1103/PhysRevC.90.054901} {\bibfield  {journal} {\bibinfo
  {journal} {Phys. Rev.}\ }\textbf {\bibinfo {volume} {C90}},\ \bibinfo {pages}
  {054901} (\bibinfo {year} {2014})},\ \Eprint {http://arxiv.org/abs/1406.2474}
  {arXiv:1406.2474 [nucl-ex]} \BibitemShut {NoStop}%
\bibitem [{\citenamefont {Adam}\ \emph {et~al.}(2016)\citenamefont {Adam} \emph
  {et~al.}}]{Adam:2016izf}%
  \BibitemOpen
  \bibfield  {author} {\bibinfo {author} {\bibfnamefont {J.}~\bibnamefont
  {Adam}} \emph {et~al.} (\bibinfo {collaboration} {ALICE}),\ }\href {\doibase
  10.1103/PhysRevLett.116.132302} {\bibfield  {journal} {\bibinfo  {journal}
  {Phys. Rev. Lett.}\ }\textbf {\bibinfo {volume} {116}},\ \bibinfo {pages}
  {132302} (\bibinfo {year} {2016})},\ \Eprint
  {http://arxiv.org/abs/1602.01119} {arXiv:1602.01119 [nucl-ex]} \BibitemShut
  {NoStop}%
\bibitem [{\citenamefont {Nie}\ \emph {et~al.}(2019)\citenamefont {Nie},
  \citenamefont {Yi}, \citenamefont {Jia},\ and\ \citenamefont
  {Ma}}]{Nie:2019swk}%
  \BibitemOpen
  \bibfield  {author} {\bibinfo {author} {\bibfnamefont {M.}~\bibnamefont
  {Nie}}, \bibinfo {author} {\bibfnamefont {L.}~\bibnamefont {Yi}}, \bibinfo
  {author} {\bibfnamefont {J.}~\bibnamefont {Jia}}, \ and\ \bibinfo {author}
  {\bibfnamefont {G.}~\bibnamefont {Ma}},\ }\href@noop {} {\  (\bibinfo {year}
  {2019})},\ \Eprint {http://arxiv.org/abs/1906.01422} {arXiv:1906.01422
  [nucl-th]} \BibitemShut {NoStop}%
\bibitem [{\citenamefont {Schenke}\ and\ \citenamefont
  {Schlichting}(2016)}]{Schenke:2016ksl}%
  \BibitemOpen
  \bibfield  {author} {\bibinfo {author} {\bibfnamefont {B.}~\bibnamefont
  {Schenke}}\ and\ \bibinfo {author} {\bibfnamefont {S.}~\bibnamefont
  {Schlichting}},\ }\href {\doibase 10.1103/PhysRevC.94.044907} {\bibfield
  {journal} {\bibinfo  {journal} {Phys. Rev.}\ }\textbf {\bibinfo {volume}
  {C94}},\ \bibinfo {pages} {044907} (\bibinfo {year} {2016})},\ \Eprint
  {http://arxiv.org/abs/1605.07158} {arXiv:1605.07158 [hep-ph]} \BibitemShut
  {NoStop}%
\bibitem [{\citenamefont {McDonald}\ \emph {et~al.}(2019)\citenamefont
  {McDonald}, \citenamefont {Jeon},\ and\ \citenamefont
  {Gale}}]{McDonald:2018wql}%
  \BibitemOpen
  \bibfield  {author} {\bibinfo {author} {\bibfnamefont {S.}~\bibnamefont
  {McDonald}}, \bibinfo {author} {\bibfnamefont {S.}~\bibnamefont {Jeon}}, \
  and\ \bibinfo {author} {\bibfnamefont {C.}~\bibnamefont {Gale}},\ }\href
  {\doibase 10.1016/j.nuclphysa.2018.08.014} {\bibfield  {journal} {\bibinfo
  {journal} {Nucl. Phys.}\ }\textbf {\bibinfo {volume} {A982}},\ \bibinfo
  {pages} {239} (\bibinfo {year} {2019})},\ \Eprint
  {http://arxiv.org/abs/1807.05409} {arXiv:1807.05409 [nucl-th]} \BibitemShut
  {NoStop}%
\bibitem [{\citenamefont {Shen}\ and\ \citenamefont
  {Schenke}(2018)}]{Shen:2017bsr}%
  \BibitemOpen
  \bibfield  {author} {\bibinfo {author} {\bibfnamefont {C.}~\bibnamefont
  {Shen}}\ and\ \bibinfo {author} {\bibfnamefont {B.}~\bibnamefont {Schenke}},\
  }\href {\doibase 10.1103/PhysRevC.97.024907} {\bibfield  {journal} {\bibinfo
  {journal} {Phys. Rev.}\ }\textbf {\bibinfo {volume} {C97}},\ \bibinfo {pages}
  {024907} (\bibinfo {year} {2018})},\ \Eprint
  {http://arxiv.org/abs/1710.00881} {arXiv:1710.00881 [nucl-th]} \BibitemShut
  {NoStop}%
\bibitem [{\citenamefont {Kurkela}\ \emph
  {et~al.}(2019{\natexlab{a}})\citenamefont {Kurkela}, \citenamefont
  {Mazeliauskas}, \citenamefont {Paquet}, \citenamefont {Schlichting},\ and\
  \citenamefont {Teaney}}]{Kurkela:2018wud}%
  \BibitemOpen
  \bibfield  {author} {\bibinfo {author} {\bibfnamefont {A.}~\bibnamefont
  {Kurkela}}, \bibinfo {author} {\bibfnamefont {A.}~\bibnamefont
  {Mazeliauskas}}, \bibinfo {author} {\bibfnamefont {J.-F.}\ \bibnamefont
  {Paquet}}, \bibinfo {author} {\bibfnamefont {S.}~\bibnamefont {Schlichting}},
  \ and\ \bibinfo {author} {\bibfnamefont {D.}~\bibnamefont {Teaney}},\ }\href
  {\doibase 10.1103/PhysRevLett.122.122302} {\bibfield  {journal} {\bibinfo
  {journal} {Phys. Rev. Lett.}\ }\textbf {\bibinfo {volume} {122}},\ \bibinfo
  {pages} {122302} (\bibinfo {year} {2019}{\natexlab{a}})},\ \Eprint
  {http://arxiv.org/abs/1805.01604} {arXiv:1805.01604 [hep-ph]} \BibitemShut
  {NoStop}%
\bibitem [{\citenamefont {Kurkela}\ \emph
  {et~al.}(2019{\natexlab{b}})\citenamefont {Kurkela}, \citenamefont
  {Mazeliauskas}, \citenamefont {Paquet}, \citenamefont {Schlichting},\ and\
  \citenamefont {Teaney}}]{Kurkela:2018vqr}%
  \BibitemOpen
  \bibfield  {author} {\bibinfo {author} {\bibfnamefont {A.}~\bibnamefont
  {Kurkela}}, \bibinfo {author} {\bibfnamefont {A.}~\bibnamefont
  {Mazeliauskas}}, \bibinfo {author} {\bibfnamefont {J.-F.}\ \bibnamefont
  {Paquet}}, \bibinfo {author} {\bibfnamefont {S.}~\bibnamefont {Schlichting}},
  \ and\ \bibinfo {author} {\bibfnamefont {D.}~\bibnamefont {Teaney}},\ }\href
  {\doibase 10.1103/PhysRevC.99.034910} {\bibfield  {journal} {\bibinfo
  {journal} {Phys. Rev.}\ }\textbf {\bibinfo {volume} {C99}},\ \bibinfo {pages}
  {034910} (\bibinfo {year} {2019}{\natexlab{b}})},\ \Eprint
  {http://arxiv.org/abs/1805.00961} {arXiv:1805.00961 [hep-ph]} \BibitemShut
  {NoStop}%
\bibitem [{\citenamefont {Oliinychenko}\ and\ \citenamefont
  {Koch}(2019)}]{Oliinychenko:2019zfk}%
  \BibitemOpen
  \bibfield  {author} {\bibinfo {author} {\bibfnamefont {D.}~\bibnamefont
  {Oliinychenko}}\ and\ \bibinfo {author} {\bibfnamefont {V.}~\bibnamefont
  {Koch}},\ }\href@noop {} {\  (\bibinfo {year} {2019})},\ \Eprint
  {http://arxiv.org/abs/1902.09775} {arXiv:1902.09775 [hep-ph]} \BibitemShut
  {NoStop}%
\bibitem [{\citenamefont {Singh}\ \emph {et~al.}(2019)\citenamefont {Singh},
  \citenamefont {Shen}, \citenamefont {McDonald}, \citenamefont {Jeon},\ and\
  \citenamefont {Gale}}]{Singh:2018dpk}%
  \BibitemOpen
  \bibfield  {author} {\bibinfo {author} {\bibfnamefont {M.}~\bibnamefont
  {Singh}}, \bibinfo {author} {\bibfnamefont {C.}~\bibnamefont {Shen}},
  \bibinfo {author} {\bibfnamefont {S.}~\bibnamefont {McDonald}}, \bibinfo
  {author} {\bibfnamefont {S.}~\bibnamefont {Jeon}}, \ and\ \bibinfo {author}
  {\bibfnamefont {C.}~\bibnamefont {Gale}},\ }\href {\doibase
  10.1016/j.nuclphysa.2018.10.061} {\bibfield  {journal} {\bibinfo  {journal}
  {Nucl. Phys.}\ }\textbf {\bibinfo {volume} {A982}},\ \bibinfo {pages} {319}
  (\bibinfo {year} {2019})},\ \Eprint {http://arxiv.org/abs/1807.05451}
  {arXiv:1807.05451 [nucl-th]} \BibitemShut {NoStop}%
\end{thebibliography}%
\end{document}